\documentstyle[preprint,aps]{revtex}

\tighten
\begin{document}

\title{
       Few-nucleon systems in translationally invariant 
       harmonic oscillator basis
}
\medskip

\author{
        P. Navr\'atil$^{a,b}$, G. P. Kamuntavi\v{c}ius$^{a,c,d}$ 
        and B. R. Barrett$^{a}$
        }

\medskip

\address{  $^a$Department of Physics, University of Arizona, 
        	Tucson, Arizona 85721\\
           $^b$Institute of Nuclear Physics, Academy of Sciences 
           	of the Czech Republic,\\
               	250 68 \v{R}e\v{z} near Prague, Czech Republic\\
           $^c$Vytautas Magnus University, Kaunas LT-3000, Lithuania\\
           $^d$Institute of Physics, Vilnius LT-2600, Lithuania
}

\maketitle

\bigskip

\begin{abstract}
We present a translationally invariant formulation of the no-core shell
model approach for few-nucleon systems. We discuss a general method 
of antisymmetrization
of the harmonic-oscillator (HO) basis depending on Jacobi coordinates.  
The use of a translationally invariant basis allows us to employ larger 
model spaces than in traditional shell-model calculations. 
Moreover, in addition to two-body effective interactions, 
three- or higher-body effective interactions as well as real three-body 
interactions can be utilized. In the present study we apply the formalism
to solve three and four nucleon systems interacting by
the CD-Bonn nucleon-nucleon (NN) potential in model spaces that include 
up to $34\hbar\Omega$ and $16\hbar\Omega$ 
HO excitations, respectively. Results of ground-state 
as well as excited-state energies, rms radii and magnetic moments 
are discussed. In addition, we compare charge form factor results
obtained using the CD-Bonn and Argonne V8' NN potentials.  
\end{abstract}

\bigskip
\bigskip
\bigskip

\narrowtext



\section{Introduction}
\label{sec1}

Various methods have been used to solve the few-nucleon problem 
in the past. The Faddeev method \cite{Fad60}
has been successfully applied to solve the three-nucleon bound-state
problem for different nucleon-nucleon potentials \cite{CPFG85,FPSS93,NHKG97}.
For the solution of the four-nucleon problem one can employ Yakubovsky's 
generalization
of the Faddeev formalism  \cite{Ya67} as done, e.g., in Refs. \cite{GH93}
or \cite{CC98}.
Alternatively, other methods have also been succesfully used in the past,
such as, 
the correlated hyperspherical harmonics expansion method \cite{VKR95,BLO99}
or the Green's function Monte Carlo method \cite{GFMC}. 
Apart from the coupled cluster method \cite{CCM,HM99} applicable typically
to closed-shell nuclei, 
the Green's function Monte Carlo method is the only approach at the present time,
for which exact solutions of systems with $A>4$ interacting by realistic 
potentials can be obtained, with the current limit being $A=8$.

On the other hand, when studying the properties of more
complex nuclei, one typically resorts to the shell model.
In that approach, the single-particle harmonic-oscillator basis 
is used and the calculations are performed in a truncated model 
space. Instead of the free 
NN potential, one utilizes effective interactions 
appropriate for the truncated model space. 
Examples of such calculations are the
large-basis no-core shell-model calculations that have recently been 
performed \cite{ZBVHS,NB96}. 
In these calculations, the effective interaction is
determined for a system of two nucleons bound in a HO well and
interacting by the nucleon-nucleon potential.
We note that the use of a HO basis is crucial for insuring that
the center-of-mass motion of the nucleus 
does not mix with the internal motion of the nucleons.
On the one hand, this approach is limited
by the model-space size, and, on the other hand,
by the fact that only a
two-body effective interaction is used despite the fact that
higher-body effective interactions might not be negligible.

It is possible, however, to re-formulate the shell-model problem
in a translationally-invariant way \cite{KM66,MM71,GK88}.
Recently, we combined the no-core shell-model approach 
to the three- and four-nucleon systems with the use of
antisymmetrized translationally invariant HO basis \cite{NB98,NB99}. 
That allowed us, due to the omission of the center of mass, 
to extend the shell-model
calculations to model spaces of $32\hbar\Omega$ and
$14\hbar\Omega$ excitations above the unperturbed ground state 
for the $A=3$ and $A=4$
systems, respectively. In addition, that approach made it possible
to employ the three-body effective interactions in the $A=4$ calculations.

In the present paper we simplify and generalize this approach so that
it is applicable
to an arbitrary number of nucleons. In particular, we discuss in detail
how to construct an antisymmetrized HO basis depending on Jacobi coordinates.
We present an iterative formula for computing the antisymmetrized
basis for $A$ nucleons from the antisymmetrized basis for $A-1$ nucleons.
Further, we discuss how to transform the antisymmetrized states
to bases containing different antisymmetrized subclusters of nucleons.

We also describe the effective interaction derivation from a different
perspective than in our previous papers. Namely, we point out the 
connections between the no-core shell-model approach 
and the unitary-model-operator
approach \cite{UMOA}. Let us remark that a fundamental feature of the
effective interactions that we employ is the fact that with the 
increasing model-space size the effective interactions approach
the bare NN interaction. Therefore, in principle,
our approach converges to the exact few-nucleon solution. 

The basic advantage of this formalism is, first, the fact that larger
model spaces can be utilized than in the standard shell-model calculations,
because the center-of-mass degrees of freedom are omitted and because
a coupled basis with good $J$ and $T$ is used. 
Second, due to the flexibility of HO
states depending on Jacobi coordinates, different recouplings of the
basis are possible. Consequently, not only can two-body 
effective interactions be utilized, but also 
three- or higher-body effective interactions as well as real three-body 
interactions. On the other hand, because the antisymmetrization
procedure is computationally involved, the practical applicability of the
formalism is limited to light nuclei. In the present formulation we expect
that significant improvement over the traditional shell-model results
can be achieved for $A\leq 6$.

We apply the formalism to solve three- and four-nucleon systems
interacting by the non-local momentum-space CD-Bonn NN 
potential \cite{Machl}. The present calculations are done in larger
model spaces than those used in Refs. \cite{NB98,NB99}. 
Also, there have not been any published results so far for the $A=4$ 
system interacting
by the CD-Bonn NN potential. In addition to the calculation of ground-state
and excited-state energies, point-nucleon rms radii and magnetic moments, 
we evaluate electromagnetic (EM) and strangeness form factors
in the impulse approximation. We plan to present the results of calculations
for $A=5$ and $A=6$, using the present formalism, separately.

In Section \ref{sec2}, we discuss the standard no-core shell-model
formulation, i.e., the Hamiltonian and effective interaction calculation.
The construction of the translationally invariant HO basis is
described in Section \ref{sec3}. Results for $A=3$ and $A=4$ systems
interacting by the CD-Bonn NN potential are given in Section \ref{sec4}.
In Section \ref{sec5}, we present concluding remarks.

\section{No-core shell-model approach}
\label{sec2}

\subsection{Hamiltonian}

In the no-core shell-model approach we start from the one- plus 
two-body Hamiltonian for the $A$-nucleon system, i.e.,
\begin{equation}\label{ham}
H_A=\sum_{i=1}^A \frac{\vec{p}_i^2}{2m}+\sum_{i<j=1}^A 
V_{\rm N}(\vec{r}_i-\vec{r}_j) \; ,
\end{equation}
where $m$ is the nucleon mass and $V_{\rm N}(\vec{r}_i-\vec{r}_j)$, 
the NN interaction. In order to simplify the notation, the spin and isospin 
dependence is omitted in the interaction term in Eq. (\ref{ham}).
We can use both coordinate-space dependent
NN potentials, such as the Reid, Nijmegen \cite{SKTS} or Argonne \cite{GFMC} 
as well as momentum-space dependent
NN potentials, such as the CD-Bonn \cite{Machl}. 
In the next step we modify the Hamiltonian (\ref{ham}) 
by adding to it the center-of-mass HO potential
$\frac{1}{2}Am\Omega^2 \vec{R}^2$, 
$\vec{R}=\frac{1}{A}\sum_{i=1}^{A}\vec{r}_i$.
This potential does not influence intrinsic properties of the 
many-body system. It allows us, however, to work with a convenient 
HO basis and provides a mean field that facilitates the calculation 
of effective interactions.
The modified Hamiltonian, depending on the HO 
frequency $\Omega$, can be cast into the form
\begin{equation}\label{hamomega}
H_A^\Omega=\sum_{i=1}^A \left[ \frac{\vec{p}_i^2}{2m}
+\frac{1}{2}m\Omega^2 \vec{r}^2_i
\right] + \sum_{i<j=1}^A \left[ V_{\rm N}(\vec{r}_i-\vec{r}_j)
-\frac{m\Omega^2}{2A}
(\vec{r}_i-\vec{r}_j)^2
\right] \; .
\end{equation}
The interaction term of the Hamiltonian (\ref{hamomega}) depends
only on the relative coordinates.
The one-body term in Eq. (\ref{hamomega}) can be re-written
as a sum of the center-of-mass term,
and a term depending on the relative coordinates.

The shell-model calculations are performed in a finite model space. 
Therefore, the interaction term in Eq. (\ref{hamomega}) must be replaced
by an effective interaction. In general, for an $A$-nucleon system,
an $A$-body
effective interaction is needed. In practice, the effective
interaction is usually approximated by a two-body effective interaction.
In the present study we will also employ a three-body effective 
interaction. 
As approximations are involved in the effective interaction
treatment, large model spaces
are desirable. In that case, the calculation
should be less affected by any imprecision of the effective
interaction. 
The same is true for the evaluation of any observable characterized
by an operator. In the model space, renormalized effective operators 
are required. The larger the model space, the less renormalization
is needed.  

As the Hamiltonian $H_A^\Omega$ (\ref{hamomega})
differs from the Hamiltonian $H_A$ (\ref{ham})
only by a center-of-mass dependent term, no dependence on $\Omega$
should exist for the intrinsic properties of the nucleus. However,
because of the approximations involved in the effective interaction
derivation, a dependence on $\Omega$ appears in our calculations. This
dependence decreases as the size of the model-space is increased.

\subsection{Effective interaction theory in Lee-Suzuki approach}
\label{LS}

In our approach we employ the Lee-Suzuki similarity transformation
method \cite{LS80,S82SO83}, which yields 
a starting-energy independent Hermitian effective interaction.
In this subsection we recapitulate general formulation and basic results
of this method. Applications of this method for computation of two- or
three-body effective interactions are described in the following
subsections. 

Let us consider an ${\it arbitrary}$ Hamiltonian $H$ with the eigensystem
$E_k, |k\rangle$, i.e.,
\begin{equation}\label{schreq}
H|k\rangle = E_k |k\rangle \; .
\end{equation}
Let us further divide the full space into the model space defined by 
a projector $P$ and the complementary space defined by a projector
$Q$, $P+Q=1$.
A similarity transformation of the Hamiltonian $e^{-\omega} H e^\omega$ 
can be introduced with a transformation operator $\omega$ satisfying 
the condition $\omega= Q \omega P$. The transformation operator
is then determined from the requirement of decoupling of the Q-space and
the model space as follows
\begin{equation}\label{decoupl}
Q e^{-\omega} H e^\omega P = 0 \; .
\end{equation}
If we denote the model space basis states as $|\alpha_P\rangle$,
and those which belong to the Q-space, as $|\alpha_Q\rangle$,
then the relation $Q e^{-\omega} H e^\omega P |k\rangle = 0$,
following from Eq. (\ref{decoupl}), will be satisfied for a particular
eigenvector $|k\rangle$ of the Hamiltonian (\ref{schreq}),
if its Q-space components can be expressed as a combination
of its P-space components with the help of the transformation operator 
$\omega$, i.e.,
\begin{equation}\label{eigomega}  
\langle\alpha_Q|k\rangle=\sum_{\alpha_P}
\langle\alpha_Q|\omega|\alpha_P\rangle \langle\alpha_P|k\rangle \; .
\end{equation}
If the dimension of the model space is $d_P$, we may choose a set
${\cal K}$ of $d_P$ eigenevectors, 
for which the relation (\ref{eigomega}) 
will be satisfied. Under the condition that the $d_P\times d_P$ 
matrix $\langle\alpha_P|k\rangle$ for $|k\rangle\in{\cal K}$
is invertible, the operator $\omega$ can be determined from 
(\ref{eigomega}) as
\begin{equation}\label{omegasol}
\langle\alpha_Q|\omega|\alpha_P\rangle = \sum_{k \in{\cal K}}
\langle\alpha_Q|k\rangle\langle\tilde{k}|\alpha_P\rangle \; ,
\end{equation}  
where we denote by tilde the inverted matrix of $\langle\alpha_P|k\rangle$, e.g.,
$\sum_{\alpha_P}\langle\tilde{k}|\alpha_P\rangle\langle\alpha_P
|k'\rangle = \delta_{k,k'}$, for $k,k'\in{\cal K}$.

The Hermitian effective Hamiltonian defined on the model space $P$
is then given by \cite{S82SO83}
\begin{equation}\label{hermeffomega}
\bar{H}_{\rm eff}
=\left[P(1+\omega^\dagger\omega)P\right]^{1/2}
PH(P+Q\omega P)\left[P(1+\omega^\dagger\omega)
P\right]^{-1/2} \; .
\end{equation}
By making use of the properties of the operator $\omega$,
the effective Hamiltonian $\bar{H}_{\rm eff}$ can be rewritten
in an explicitly Hermitian form as
\begin{equation}\label{exhermeff}
\bar{H}_{\rm eff}
=\left[P(1+\omega^\dagger\omega)P\right]^{-1/2}
(P+P\omega^\dagger Q)H(Q\omega P+P)\left[P(1+\omega^\dagger\omega)
P\right]^{-1/2} \; .
\end{equation}
With the help of the solution for $\omega$ (\ref{omegasol})
we obtain a simple expression for the matrix elements of 
the effective Hamiltonian
\begin{eqnarray}\label{effham}
\langle \alpha_P | \bar{H}_{\rm eff} |\alpha_{P'}\rangle
&=& \sum_{k \in{\cal K}}\sum_{\alpha_{P''}}\sum_{\alpha_{P'''}} 
\langle \alpha_P | (1+\omega^\dagger\omega)^{-1/2}
|\alpha_{P''}\rangle
\langle \alpha_{P''} |\tilde{k}\rangle E_k 
\langle\tilde{k}|\alpha_{P'''}\rangle
\nonumber \\
&&\times
\langle \alpha_{P'''} | (1+\omega^\dagger\omega)^{-1/2}
|\alpha_{P'}\rangle \; .
\end{eqnarray}
For computation of the matrix elements of 
$(1+\omega^\dagger\omega)^{-1/2}$, we can use the relation
\begin{equation}\label{omdegom}
\langle \alpha_P | (1+\omega^\dagger\omega)
|\alpha_{P''}\rangle = \sum_{k \in{\cal K}} \langle \alpha_P |
\tilde{k} \rangle \langle \tilde{k}|\alpha_{P''}\rangle \; ,
\end{equation}
to remove the summation over the Q-space basis states.
The effective Hamiltonian (\ref{effham}) reproduces the
eigenenergies $E_k, k\in{\cal K}$ in the model space.
We note that the relation (\ref{effham}) used together with 
(\ref{omdegom}) does not contain any summation over the Q-space 
basis and, thus, represents a simplification compared to the formulas
we presented in our previous papers \cite{NB96,NB98,NB99},
though it is fully equivalent.

\subsection{Unitary transformation of the Hamiltonian 
and the two-body effective interaction}

We now return to our problem, namely, to derive the effective
interaction corresponding to a chosen model space
for the {\it particular} Hamiltonian $H_A^\Omega$ (\ref{hamomega}).
Let us write the Hamiltonian (\ref{hamomega}) schematically as
\begin{equation}\label{UMOAstham}
H_A^\Omega = \sum_{i=1}^{A} h_i + \sum_{i<j=1}^{A} V_{ij} \; .
\end{equation}
According to Providencia and Shakin \cite{PS64}, a unitary transformation
of the Hamiltonian, which is able to accomodate the short-range two-body 
correlations in a nucleus, 
can be introduced by choosing a two-body antihermitian operator $S$,
such that
\begin{equation}\label{UMOAtrans}
{\cal H} = e^{-S} H_A^\Omega e^{S} \; .
\end{equation}
Consequently, the transformed Hamiltonian can be expanded in a cluster
expansion
\begin{equation}\label{UMOAexpan}
{\cal H} = {\cal H}^{(1)} + {\cal H}^{(2)} + {\cal H}^{(3)} +\ldots \; ,
\end{equation}
where the one-body, two-body and three-body terms are given as
\begin{mathletters}\label{UMOAterms}
\begin{eqnarray}
{\cal H}^{(1)} &=& \sum_{i=1}^{A} h_i \; , \\
{\cal H}^{(2)} &=& \sum_{i<j=1}^{A} \tilde{V}_{ij} \; , \\
{\cal H}^{(3)} &=& \sum_{i<j<k=1}^{A} \tilde{V}_{ijk} \; ,
\end{eqnarray}
\end{mathletters}
with
\begin{mathletters}\label{UMOAexplterms}
\begin{eqnarray}
\tilde{V}_{12} &=& e^{-S_{12}}(h_1+h_2+V_{12})e^{S_{12}}-(h_1+h_2)   
\; , \label{V12eff} \\
\tilde{V}_{123} &=& e^{-S_{123}}(h_1+h_2+h_3+V_{12}+V_{13}+V_{23})
e^{S_{123}} 
\nonumber \\
&&-(h_1+h_2+h_3+\tilde{V}_{12}+\tilde{V}_{13}+\tilde{V}_{23})   
\; ,
\end{eqnarray}
\end{mathletters}
and $S_{123}=S_{12}+S_{23}+S_{31}$. In the above equations, it has been
assumed that the basis states are eigenstates of the one-body, in 
our case HO, Hamiltonian $\sum_{i=1}^A h_i$.

If the full space is divided into a model space and a Q-space, using
the projectors $P$ and $Q$ with $P+Q=1$, 
it is possible to determine the transformation operator $S_{12}$
from the decoupling condition 
\begin{equation}\label{UMOAdecoupl}
Q_{2} e^{-S_{12}}(h_1+h_2+V_{12})e^{S_{12}} P_{2} = 0 \; ,
\end{equation}
and the simultaneous restrictions $P_2 S_{12} P_2 = Q_2 S_{12} Q_2 =0$. 
Note that two-nucleon-state projectors
appear in Eq. (\ref{UMOAdecoupl}), whose definitions follow from the
definitions of the $A$-nucleon projectors $P$, $Q$.
This approach, introduced by Suzuki and Okamoto and 
referred to as the unitary-model-operator approach (UMOA) \cite{UMOA},
has a solution that can be expressed in the following form
\begin{equation}\label{UMOAsol}
S_{12} = {\rm arctanh}(\omega-\omega^\dagger)    \; ,
\end{equation}
with the operator $\omega$ satisfying $\omega=Q_2\omega P_2$. This is the same
operator that solves Eq. (\ref{decoupl}) with $H=h_1+h_2+V_{12}$.
Moreover, $P_{2} e^{-S_{12}}(h_1+h_2+V_{12})e^{S_{12}} P_{2}$ is given
by Eq. (\ref{exhermeff}) with $H=h_1+h_2+V_{12}$.

\subsection{Two-body effective interaction calculation}
\label{v2beff}

We compute the two-body effective interaction according to
Eq. (\ref{V12eff}) using the decoupling condition (\ref{UMOAdecoupl}).
For the two-nucleon Hamiltonian we employ
\begin{equation}\label{hamomega2}
H_2^\Omega = H_{02}+V_{12}=
\frac{\vec{p}^2}{2m}
+\frac{1}{2}m\Omega^2 \vec{r}^2
+ V_{\rm N}(\sqrt{2}\vec{r})-\frac{m\Omega^2}{A}\vec{r}^2 \; ,
\end{equation}
where $\vec{r}=\sqrt{\frac{1}{2}}(\vec{r}_1-\vec{r}_2)$ and
$\vec{p}=\sqrt{\frac{1}{2}}(\vec{p}_1-\vec{p}_2)$ and
where $H_{02}$ differs from $h_1+h_2$ by the omission of 
the center-of-mass HO term of nucleons 1 and 2.
Our calculations start with exact solutions of the Hamiltonian
(\ref{hamomega2}) and, consequently, we construct
the operator $\omega$ and, then, the effective 
interaction directly from these solutions by application of 
the relations (\ref{omegasol}) and (\ref{effham}) with $E_k, |k\rangle$
obtained from the solution of the Schr\"odinger equation (\ref{schreq}) 
for the Hamiltonian $H = H_2^\Omega$. 
The relative-coordinate two-nucleon 
HO states used in the calculation are characterized by
quantum numbers $|nlsjt\rangle$ with the radial and orbital
HO quantum numbers corresponding to coordinate $\vec{r}$ 
and momentum $\vec{p}$. Typically, we solve the two-nucleon
Hamiltonian (\ref{hamomega2}) for all two-nucleon channels up
to $j=6$. For the channels with higher $j$ only the kinetic-energy 
term is used in the many-nucleon calculation.
The model space is defined by the maximal number of allowed HO 
excitations $N_{\max}$ from the condition $2n+l\leq N_{\rm max}$.

In order to construct the operator $\omega$ (\ref{omegasol})
we need to select the set of eigenvectors ${\cal K}$.
In the present application we select the lowest states
obtained in each channel. It turns out that these states also have
the largest overlap with the model space. Their number is given 
by the number of basis states satisfying $2n+l\leq N_{\rm max}$.

Finally, the two-body effective interaction
is determined from the two-nucleon effective Hamiltonian, obtained from
(\ref{effham}), as $V_{\rm 2eff}=\bar{H}_{\rm 2eff}-H_{02}$.
Apart from being a function of the nucleon number $A$, $V_{\rm 2eff}$
depends on the HO frequency $\Omega$ and on the model-space
defining parameter $N_{\rm max}$. 
It has the important property that $V_{\rm 2eff}\rightarrow V_{12}$ 
for $N_{\rm max}\rightarrow \infty$ following from the fact that
$\omega\rightarrow 0$ for $P\rightarrow 1$.

\subsection{Three-body effective interaction calculation}
\label{v3beff}

In the standard shell-model calculations the effective interaction
is limited to a two-body effective interaction. In our approach 
the limitation to a two-body effective interaction
means that we use $V_{\rm 2eff}$, computed as discussed
in the previous subsection, and neglect all the three- and higher-body
clusters appearing in the expansion (\ref{UMOAexpan}). The formalism
laid out in the present paper allows us, however, 
to employ three-body or even
higher-body interactions in a straightforward manner. Therefore,
we are in a position to go beyond the two-body effective interaction
approximation. One way to procceed is to evaluate the three-body
term appearing in 
Eqs. (\ref{UMOAexpan},\ref{UMOAterms},\ref{UMOAexplterms}).
That term depends, through $S_{123}$ in Eq. (\ref{UMOAexplterms}), 
on the operator $\omega$, appearing in Eq. (\ref{UMOAsol}), 
used for the construction of the two-body effective interaction.  
It does not, however, guarantee the model space and the Q-space
decoupling for the three-nucleon system in a similar way as 
in Eq. (\ref{UMOAdecoupl}) for the two-nucleon system. To improve
on this point, we, therefore, choose a different way of deriving
the three-body effective interaction. 
In our previous paper \cite{NB99} we introduced a procedure 
applicable for nuclei with $A\geq 4$ that
calculates the three-body correlations directly and guarantees
the model space and Q-space decoupling for the three-nucleon system.

In order to calculate the three-body effective interaction in this way, 
we first re-write the two-body interaction term of the Hamiltonian
(\ref{hamomega},\ref{UMOAstham}) using
\begin{equation}\label{twothree}
\sum_{i<j=1}^A V_{ij}=\frac{1}{A-2}
\sum_{i<j<k=1}^A (V_{ij}+V_{ik}+V_{jk}) 
\; ,
\end{equation} 
valid for $A\geq 3$. Then, formally, our Hamiltonian consists only
of one-body and three-body terms. We now calculate the three-body
effective interaction that corresponds to $V_{ij}+V_{ik}+V_{jk}$
from the three-nucleon system condition
\begin{equation}\label{UMOA3decoupl}
Q_{3} e^{-S^{(3)}}(h_1+h_2+h_3+V_{12}+V_{13}+V_{23})e^{S^{(3)}} 
P_{3} = 0 \; ,
\end{equation}
in complete analogy to Eq. (\ref{UMOAdecoupl}). Note that $S^{(3)}$ 
is different from $S_{123}$ and is determined by (\ref{UMOA3decoupl}). 
The three-body effective interaction
is then obtained utilizing the solutions of the three-nucleon system 
solutions from
Eqs. (\ref{schreq},\ref{omegasol},\ref{effham}) for the Hamiltonian
\begin{equation}\label{hamomega3}
H_3^\Omega = h_1+h_2+h_3+V_{12}+V_{13}+V_{23} \; .
\end{equation}
As the interaction depends only on the relative positions of nucleons
1, 2 and 3, the three-nucleon center of mass can be separated,
when solving
Schr\"{o}dinger equation with $H_3^\Omega$. Similarly as for the 
two-nucleon Hamiltonian (\ref{hamomega2}), the center-of-mass term
is not considered in the effective-interaction calculation. We obtain
the three-nucleon solutions corresponding to the Hamiltonian 
(\ref{hamomega3}) by, first, introducing Jacobi coordinates, as described 
later in this paper and also as described in our previous papers 
\cite{NB98,NB99,NBG99} and, by, second, the introducing for 
the interactions $V_{12},V_{13},V_{23}$ the two-body effective interactions
corresponding 
to a large space characterized by $N_{\rm 3max}\approx 30$ and derived
according to the procedure described in the previous subsection.
A space of this size is sufficient for obtaining exact or almost
exact solutions of the three-nucleon problem \cite{NB98}. 
We note that
\begin{equation}\label{V12}
V_{ij}=V_{\rm N}(\vec{r}_i-\vec{r}_j)-\frac{m\Omega^2}{2A}
(\vec{r}_i-\vec{r}_j)^2 \; ,  \; \; i,j=1,2,3 \; ,
\end{equation}
yields a three-nucleon bound system for $A\geq 4$. 
Therefore, as in the case of the two-body effective interaction
calculation, the Lee-Suzuki approach is applicable, as described
in Subsection \ref{LS}, in a straightforward way. We use the solutions
of the three-nucleon system to construct the operator $\omega$ 
and, then, the three-body effective 
interaction directly from these solutions by applications of 
the relations (\ref{omegasol}) and (\ref{effham}). The eigensystem 
$E_k, |k\rangle$ is 
obtained from the solution of Schr\"odinger equation (\ref{schreq}) 
for the Hamiltonian (\ref{hamomega3}) with the interactions $V_{ij}$
replaced by $V_{{\rm 2eff},ij}$, where the two-body effective interaction
corresponds to the space defined by $N_{\rm 3max}$, as discussed above.
The model space, defined by the projector $P_3$, is characterized
by the maximal allowed number of HO excitations $N_{\rm max}$, 
where $N_{\rm max}<N_{\rm 3max}$. It is spanned by all states satisfying the
condition $N_3\leq N_{\rm max}$ with 
$N_3$ given as $N_3=2n+l+2{\cal N}+{\cal L}$ when basis (\ref{hobas})
is used, or, using the basis (\ref{abas3}) $N_3\equiv N$.
The Q-space defined by $Q_3$ is spanned by states with total number 
of HO excitations 
$N_{\rm max}<N_3\leq N_{\rm 3max}$. Typically, we solve the three-nucleon
system and construct the three-body effective interaction
only for three-nucleon channels with 
$J_3=1/2^{\pm},3/2^\pm$ and $T_3=1/2$. For all other channels the two-body
effective interaction $\sum_{i<j=1}^{3}V_{{\rm 2eff},ij}$
corresponding to the model space characterized 
by $N_{\rm max}$ is used instead.

In order to construct the operator $\omega$ (\ref{omegasol})
we need to select the set of eigenvectors ${\cal K}$.
As in the two-body effective interaction calculation, 
the lowest states obtained in each channel are selected. 
Again, those states also have
the largest overlap with the model space. Their number is given 
by the number of basis states with the total number of HO 
excitations $N_3\leq N_{\rm max}$.
The three-body effective interaction
is determined from the three-nucleon effective Hamiltonian, obtained from
(\ref{effham}), as $V_{\rm 3eff}=\bar{H}_{\rm 3eff}-H_{03}$
with $H_{03}$ equal to $h_1+h_2+h_3$ minus the three-nucleon 
center-of-mass HO term.
Like the two-body effective interaction $V_{\rm 2eff}$, 
apart from being a function of the nucleon number $A$, $V_{\rm 3eff}$
depends on the HO frequency $\Omega$ and on the model-space
defining-parameter $N_{\rm max}$. In addition, it also depends
on the choice of $N_{\rm 3max}$. Obviously, $N_{\rm 3max}$ must
be sufficiently large, in order to make this dependence negligible.
The limiting properties of $V_{\rm 3eff}$ are as follows:
$V_{\rm 3eff}\rightarrow \sum_{i<j=1}^{3}V_{{\rm 2eff},ij}$ 
for $N_{\rm max}\rightarrow N_{\rm 3max}$ and
$V_{\rm 3eff}\rightarrow \sum_{i<j=1}^{3}V_{ij}$ 
for $N_{\rm max}, N_{\rm 3max}\rightarrow \infty$.

\section{Translationally-invariant harmonic-oscillator basis}
\label{sec3}

As discussed in the previous sections, by using the effective 
interaction theory we arrive at a Hamiltonian that has 
the following structure
\begin{equation}\label{hameff}
H_{A\rm eff}^\Omega=\sum_{i=1}^A \left[ \frac{\vec{p}_i^2}{2m}
+\frac{1}{2}m\Omega^2 \vec{r}^2_i\right] 
+ \left\{\sum_{i<j=1}^A \left[ V_{\rm N}(\vec{r}_i-\vec{r}_j)
-\frac{m\Omega^2}{2A}(\vec{r}_i-\vec{r}_j)^2\right]
\right\}_{\rm eff} \; ,
\end{equation}
with the interaction term depending on relative coordinates 
(and/or relative momenta) only.
The center-of-mass dependence appears only in the one-body HO term.
Consequently, by performing a transformation to Jacobi coordinates,
the center-of-mass degrees of freedom can be removed.

\subsection{Jacobi coordinates}

We work in the isospin formalism and consider nucleons with the mass
$m$. A generalization to the proton-neutron formalism with
unequal masses for the proton and the neutron is straightforward.
We will use Jacobi coordinates that are introduced as an orthogonal
transformation of the single-nucleon coordinates. In general,
Jacobi coordinates are proportional to differences of centers of mass
of nucleon subclusters. 

For our purposes we need three different sets
of Jacobi coordinates. The first set, i.e.,   
\begin{mathletters}\label{jacobiam11}
\begin{eqnarray}
\vec{\xi}_0 &=& \sqrt{\frac{1}{A}}\left[\vec{r}_1+\vec{r}_2
                                   +\ldots +\vec{r}_A\right]
\; , \\
\vec{\xi}_1 &=& \sqrt{\frac{1}{2}}\left[\vec{r}_1-\vec{r}_2
                                                     \right]
\; , \\
\vec{\xi}_2 &=& \sqrt{\frac{2}{3}}\left[\frac{1}{2}
                 \left(\vec{r}_1+\vec{r}_2\right)
                                   -\vec{r}_3\right]
\; , \\
&\ldots & \nonumber
\\
\vec{\xi}_{A-2} &=& \sqrt{\frac{A-2}{A-1}}\left[\frac{1}{A-2}
      \left(\vec{r}_1+\vec{r}_2 + \ldots+ \vec{r}_{A-2}\right)
                                   -\vec{r}_{A-1}\right]
\; , \\
\vec{\xi}_{A-1} &=& \sqrt{\frac{A-1}{A}}\left[\frac{1}{A-1}
      \left(\vec{r}_1+\vec{r}_2 + \ldots+ \vec{r}_{A-1}\right)
                                   -\vec{r}_{A}\right]
\; ,
\end{eqnarray}
\end{mathletters}
is used for the construction of the antisymmetrized HO basis.
Here, $\vec{\xi}_0$ is proportional to the center of mass of the
$A$-nucleon system. On the other hand, $\vec{\xi}_\rho$ is proportional
to the relative position of the $\rho+1$-st nucleon and the
center of mass of the $\rho$ nucleons.  

Another set that is suitable for the basis expansion, when
two-body interaction matrix elements need to be calculated, 
is obtained by keeping 
$\vec{\xi}_0,\vec{\xi}_{1},\ldots \vec{\xi}_{A-3}$
and introducing two different variables, as follows
\begin{mathletters}\label{jacobiam22}
\begin{eqnarray}
\vec{\xi}_0,\vec{\xi}_{1},&\ldots &\vec{\xi}_{A-3}
\; , \\
\vec{\eta}_{A-2} &=& \sqrt{\frac{2(A-2)}{A}}\left[\frac{1}{A-2}
      \left(\vec{r}_1+\vec{r}_2 + \ldots+ \vec{r}_{A-2}\right)
             -\frac{1}{2}(\vec{r}_{A-1}+\vec{r}_{A})\right]
\; , \\
\vec{\eta}_{A-1} &=& \sqrt{\frac{1}{2}}\left[\vec{r}_{A-1}-\vec{r}_{A}
                                                     \right]
\; .
\end{eqnarray}
\end{mathletters}

Eventually, a set suitable for the basis expansion, when
three-body interaction matrix elements need to be calculated, 
is obtained by keeping
$\vec{\xi}_0,\vec{\xi}_{1},\ldots \vec{\xi}_{A-4}$ and
$\vec{\eta}_{A-1}$ from the previous set and introducing
other two different variables
\begin{mathletters}\label{jacobiam33}
\begin{eqnarray}
\vec{\xi}_0,\vec{\xi}_{1},&\ldots&\vec{\xi}_{A-4}
\; , \\
\vec{\vartheta}_{A-3} &=& \sqrt{\frac{3(A-3)}{A-2}}\left[\frac{1}{A-3}
      \left(\vec{r}_1+\vec{r}_2 + \ldots+ \vec{r}_{A-3}\right)
        -\frac{1}{3}(\vec{r}_{A-2}+\vec{r}_{A-1}+\vec{r}_{A})\right]
\; , \\
\vec{\vartheta}_{A-2} &=& \sqrt{\frac{2}{3}}\left[\frac{1}{2}
                 \left(\vec{r}_{A-1}+\vec{r}_A\right)
                                   -\vec{r}_{A-2}\right]
\; , \\
\vec{\eta}_{A-1} &=& \sqrt{\frac{1}{2}}\left[\vec{r}_{A-1}-\vec{r}_{A}
                                                     \right]
\; .
\end{eqnarray}
\end{mathletters}
Identical transformations, as in 
Eqs. (\ref{jacobiam11},\ref{jacobiam22},\ref{jacobiam33}),
are also introduced for the momenta.

Let us note that the one-body HO potential in Eq. (\ref{hameff})
transforms as
\begin{equation}\label{u0trans}
\sum_{i=1}^A \frac{1}{2}m\Omega^2 \vec{r}^2_i = 
\frac{1}{2}m\Omega^2 \vec{\xi}_0^2 +
\sum_{\rho=1}^{A-1} \frac{1}{2}m\Omega^2 \vec{\xi}^2_\rho \; ,
\end{equation}
and the kinetic term transforms in an analogous way. As the interaction
does not depend on $\vec{\xi}_0$ or the center-of-mass momentum,
the center-of-mass HO term can be omitted. Moreover, we can use
the HO basis, depending on coordinates 
$\vec{\xi}_\rho, \rho=1,2\ldots A-1$, e.g.,
\begin{equation}\label{jacobibas}
\prod_{\rho=1}^{A-1} \langle \vec{\xi}_\rho |n_\rho l_\rho \rangle \; ,
\end{equation}
for our calculations. First, let us remark that due to the othogonality
of the transformations 
(\ref{jacobiam11},\ref{jacobiam22},\ref{jacobiam33}),
the same HO parameter is used for all HO wave functions. Second,
we may use any of the sets
(\ref{jacobiam11},\ref{jacobiam22},\ref{jacobiam33})
for the basis construction of the type (\ref{jacobibas}),
as similar relations like (\ref{u0trans}) hold for
all the sets. Third, the basis (\ref{jacobibas}) is {\it not}
antisymmetrized with respect to the exchanges of all
nucleon pairs. The antisymmetrization procedure will be discussed
in the following subsections.

\subsection{Antisymmetrization for the three-nucleon system}

We discussed the antisymmetrization of the translationally-invariant
HO basis for the three-nucleon
system in our previous papers \cite{NB98,NB99,NBG99}. One starts 
by introducing a  
basis following from Eq. (\ref{jacobibas}) that depends on Jacobi 
coordinates
$\vec{\xi}_1$ and $\vec{\xi}_2$, defined in Eqs. (\ref{jacobiam11}),
e.g.,
\begin{equation}\label{hobas}
|(n l s j t; {\cal N} {\cal L} {\cal J}) J T \rangle \; .
\end{equation}
Here $n, l$ and ${\cal N}, {\cal L}$ are the HO quantum numbers
corresponding to the harmonic oscillators associated with the coordinates 
(and the corresponding momenta) $\vec{\xi}_1$ and $\vec{\xi}_2$, respectively. 
The quantum numbers $s,t,j$ describe the spin, isospin and angular momentum
of the relative-coordinate two-nucleon channel of nucleons 1 and 2, while 
${\cal J}$ is the angular momentum of the third nucleon relative to the
center of mass of nucleons 1 and 2. The $J$ and $T$ are the total angular 
momentum and the total isospin, respectively.
Note that the basis (\ref{hobas}) is antisymmetrized with respect
to the exchanges of nucleons 1 and 2, as the two-nucleon channel
quantum numbers are restricted by the condition $(-1)^{l+s+t}=-1$.
It is not, however, antisymmetrized with respect to the exchanges of nucleons
$1\leftrightarrow 3$ and $2\leftrightarrow 3$.
In order to construct a completely antisymmetrized basis, one needs to 
obtain eigenvectors of the antisymmetrizer 
\begin{equation}\label{antisymm3}
{\cal X}=\frac{1}{3}(1+{\cal T}^{(-)}+{\cal T}^{(+)}) \; ,
\end{equation}
where ${\cal T}^{(+)}$ and ${\cal T}^{(-)}$ are the cyclic and the anticyclic 
permutation operators, respectively. The antisymmetrizer ${\cal X}$
is a projector satisfying ${\cal X} {\cal X}={\cal X}$. 
When diagonalized in the basis (\ref{hobas}), its eigenvectors
span two eigenspaces. One, corresponding to the eigenvalue 1, is formed
by physical, completely antisymmetrized states and the other, corresponding
to the eigenvalue 0, is formed by spurious states. There are about 
twice as many spurious states as the physical ones \cite{NBG99}.

Due to the antisymmetry with respect to the exchanges $1\leftrightarrow 2$,
the matrix elements in the basis (\ref{hobas}) of the antisymmetrizer 
${\cal X}$ can be evaluated simply as 
$\langle {\cal X} \rangle = \frac{1}{3}(1-2\langle {\cal T}_{23}\rangle)$,
where ${\cal T}_{23}$ is the permutation corresponding to the exchange
of nucleons 2 and 3.
Its matrix element can be evaluated in a straightforward way, e.g.,
\begin{eqnarray}\label{t13t23}
&&\langle (n_1 l_1 s_1 j_1 t_1; {\cal N}_1 {\cal L}_1  
{\cal J}_1) J T | {\cal T}_{23} |  
(n_2 l_2 s_2 j_2 t_2; {\cal N}_2 {\cal L}_2  
{\cal J}_2) J T\rangle = \delta_{N_1,N_2} \hat{t}_1 \hat{t}_2
\left\{ \begin{array}{ccc} \textstyle{\frac{1}{2}} & \textstyle{\frac{1}{2}} 
               & t_1 \\
             \textstyle{\frac{1}{2}}  &  T  & t_2
\end{array}\right\}
\nonumber \\
&& \times \sum_{LS} \hat{L}^2 \hat{S}^2
\hat{j}_1 \hat{j}_2 \hat{\cal J}_1 \hat{\cal J}_2 \hat{s}_1 \hat{s}_2
 (-1)^L
          \left\{ \begin{array}{ccc} l_1   & s_1   & j_1   \\
          {\cal L}_1  & \textstyle{\frac{1}{2}}  & {\cal J}_1 \\
                                     L   & S  & J
\end{array}\right\}
          \left\{ \begin{array}{ccc} l_2   & s_2   & j_2   \\
       {\cal L}_2  & \textstyle{\frac{1}{2}}   & {\cal J}_2 \\
                                     L   & S  & J
\end{array}\right\}
\left\{ \begin{array}{ccc} \textstyle{\frac{1}{2}} & \textstyle{\frac{1}{2}} 
               & s_1 \\
             \textstyle{\frac{1}{2}}  &  S  & s_2
\end{array}\right\}
\nonumber \\
&&\times 
\langle n_1 l_1 {\cal N}_1 {\cal L}_1 L 
| {\cal N}_2 {\cal L}_2 n_2 l_2 L \rangle_{\rm 3} \; ,
\end{eqnarray}
where $N_i=2n_i+l_i+2{\cal N}_i+{\cal L}_i, i= 1,2$; 
$\hat{j}=\sqrt{2j+1}$; 
and $\langle n_1 l_1 {\cal N}_1 {\cal L}_1  L 
| {\cal N}_2 {\cal L}_2 n_2 l_2 L \rangle_{\rm 3}$
is the general HO bracket for two particles with mass 
ratio 3 as defined, e.g.,
in Ref. \cite{Tr72}.
The expression (\ref{t13t23}) can be derived by examining 
the action
of ${\cal T}_{23}$ on the basis states (\ref{hobas}).
That operator changes the state 
$|nl(\vec{\xi}_1),{\cal NL}(\vec{\xi}_2), L\rangle$
to $|nl(\vec{\xi'}_1),{\cal NL}(\vec{\xi'}_2), L\rangle$,
where $\vec{\xi'}_i, i=1,2$ are defined as $\vec{\xi}_i, i=1,2$
but with the single-nucleon indexes 2 and 3 exchanged. The primed Jacobi 
coordinates can be expressed as an orthogonal transformation
of the unprimed ones. Consequently, the HO wave functions
depending on the primed Jacobi coordinates can be expressed
as an orthogonal transformation of the original HO wave functions.
Elements of the transformation are the Talmi-Moshinsky HO brackets
for two particles with the mass ratio $d$, with $d$ determined
from the orthogonal transformation of the coordinates, see, e.g.,
Ref. \cite{Tr72}.

The resulting antisymmetrized states can be classified 
and expanded in terms of the original basis (\ref{hobas}) as follows
\begin{equation}\label{abas3}
|N i J T\rangle = \sum \langle nlsjt; {\cal NLJ}||N i J T\rangle 
|(nlsjt;{\cal NLJ}) JT\rangle \; ,
\end{equation}
where $N=2n+l+2{\cal N}+{\cal L}$ and 
where we introduced an
additional quantum number $i$ that distinguishes states with
the same set of quantum numbers $N, J, T$, e.g.,
$i=1,2, \ldots r$ with 
$r$ the total number of antisymmetrized states for a given $N, J, T$.
It can be obtained from computing the trace of the antisymmetrizer
${\cal X}$ \cite{GPK}
\begin{equation}\label{trace}
r= {\rm Tr}{\cal X}^{NJT} \; .
\end{equation}

\subsection{Antisymmetrization for the $A$-nucleon system}

The construction of a translationally-invariant antisymmetrized HO basis
for four nucleons that contains an antisymmetrized three-nucleon
subcluster was described in our earlier papers \cite{NB99,NBG99}. 
In this subsection
we generalize and simplify this construction to the case of an 
arbitrary number 
of nucleons $A$. The starting point is a basis that contains 
an antisymmetrized subcluster of $A-1$ nucleons, e.g.,
\begin{equation}\label{abasam11}
|(N_{\rm A-1} i_{\rm A-1} J_{\rm A-1} T_{\rm A-1};
n_{\rm A-1} l_{\rm A-1} {\cal J}_{\rm A-1})
J T\rangle \; ,
\end{equation}
with the $(A-1)$-nucleon antisymmetrized state 
$|N_{\rm A-1} i_{\rm A-1} J_{\rm A-1} T_{\rm A-1}\rangle$,
depending on the Jacobi coordinates 
$\vec{\xi}_1,\vec{\xi}_{2},\ldots \vec{\xi}_{A-2}$ (\ref{jacobiam11}),
and the state $|n_{\rm A-1} l_{\rm A-1} {\cal J}_{\rm A-1}\rangle$ 
that represents the last nucleon depending on the
Jacobi coordinate $\vec{\xi}_{A-1}$. 
For a four nucleon system
the state $|N_{\rm A-1} i_{\rm A-1} J_{\rm A-1} T_{\rm A-1}\rangle$ 
is identical to the state introduced in Eq. (\ref{abas3}). 
The basis (\ref{abasam11})
is not antisymmetrized with respect to exchanges of the last nucleon
with the others. However, due to the antisymmetry of the $A-1$ nucleon
subcluster, the matrix elements of the antisymmetrizer ${\cal X}$
in the basis (\ref{abasam11}) simplifies dramatically, e.g.,
\begin{equation}\label{xam11}
\langle {\cal X} \rangle = \frac{1}{A}\left[1-(A-1)
\langle {\cal T}_{A,A-1}\rangle\right]
\end{equation}
with ${\cal T}_{A,A-1}$ the transposition operator of the
$A$-th and the $A-1$-st nucleon. Its matrix element can be computed
in a straightforward way as
\begin{eqnarray}\label{taam1}
&&\langle (N_{\rm A-1 L} i_{\rm A-1 L} J_{\rm A-1L} T_{\rm A-1L};n_{\rm A-1 L}
l_{\rm A-1 L} {\cal J}_{\rm A-1 L}) JT | 
\nonumber \\
&&\times
{\cal T}_{\rm A,A-1}
|(N_{\rm A-1 R} i_{\rm A-1 R} J_{\rm A-1R} T_{\rm A-1R}; n_{\rm A-1 R}
l_{\rm A-1 R} {\cal J}_{\rm A-1 R}) JT \rangle \nonumber \\
&&=\delta_{N_{\rm L},N_{\rm R}} \sum
\langle{N_{\rm A-2} i_{\rm A-2} J_{\rm A-2} T_{\rm A-2};\;
n_{\rm A-2 L} l_{\rm A-2 L}{\cal J}_{\rm A-2 L}}||
{N_{\rm A-1 L} i_{\rm A-1 L} J_{\rm A-1L} T_{\rm A-1L}}\rangle
\nonumber \\
&&\times
\langle{N_{\rm A-2} i_{\rm A-2} J_{\rm A-2} T_{\rm A-2};\;
n_{\rm A-2 R} l_{\rm A-2 R}{\cal J}_{\rm A-2 R}}||
{N_{\rm A-1 R} i_{\rm A-1 R} J_{\rm A-1R} T_{\rm A-1R}}\rangle
\nonumber \\
&&\times
\hat{T}_{\rm A-1L}\hat{T}_{\rm A-1R} 
(-1)^{T_{\rm A-1L}+T_{\rm A-1R}+{\cal J}_{\rm A-2 L}+{\cal J}_{\rm A-2 R}}
\left\{ \begin{array}{ccc} \textstyle{\frac{1}{2}} & T_{\rm A-2}
               & T_{\rm A-1R} \\
             \textstyle{\frac{1}{2}}  &  T  & T_{\rm A-1L}
\end{array}\right\}
\nonumber \\
&&\times
\hat{\cal J}_{\rm A-2 L} 
\hat{\cal J}_{\rm A-2 R} \hat{\cal J}_{\rm A-1 L} 
\hat{\cal J}_{\rm A-1 R}\hat{J}_{\rm A-1L}\hat{J}_{\rm A-1R}\hat{K}^2
\left\{ \begin{array}{ccc} J_{\rm A-2} & {\cal J}_{\rm A-2 L}& J_{\rm A-1 L}  \\
  {\cal J}_{\rm A-2 R}  & K  & {\cal J}_{\rm A-1 L} \\
                          J_{\rm A-1R}   & {\cal J}_{\rm A-1 R}  & J
\end{array}\right\}
\nonumber \\
&&\times 
\left\{ \begin{array}{ccc} l_{\rm A-2 L}   & l_{\rm A-1 R}   & K   \\
  {\cal J}_{\rm A-1 R}  & {\cal J}_{\rm A-2 L}& \textstyle{\frac{1}{2}}   
\end{array}\right\}
\left\{ \begin{array}{ccc} l_{\rm A-2 R}   & l_{\rm A-1 L}   & K   \\
  {\cal J}_{\rm A-1 L}  & {\cal J}_{\rm A-2 R}& \textstyle{\frac{1}{2}}   
\end{array}\right\}
\left\{ \begin{array}{ccc} l_{\rm A-1 L}   & l_{\rm A-2 R}   &  K   \\
                           l_{\rm A-1 R}   & l_{\rm A-2 L}   &  L
\end{array}\right\}
\nonumber \\
&&\times 
\hat{L}^2 (-1)^{l_{\rm A-2 R}+l_{\rm A-1 L}+L}
\langle n_{\rm A-1 L} l_{\rm A-1 L}  n_{\rm A-2 L} l_{\rm A-2 L} \;L 
| n_{\rm A-2 R} l_{\rm A-2 R} n_{\rm A-1 R} l_{\rm A-1 R}  \;L
\rangle_{\rm A(A-2)}  \; , 
\end{eqnarray}
where 
$N_{\rm X}=N_{\rm A-1 X}+2n_{\rm A-1 X}+l_{\rm A-1 X}$, 
${\rm X}={\rm L,R}$ and 
$\langle n_{\rm A-1 L} l_{\rm A-1 L}  n_{\rm A-2 L} l_{\rm A-2 L} L 
| n_{\rm A-2 R} l_{\rm A-2 R} n_{\rm A-1 R} l_{\rm A-1 R}  L
\rangle_{\rm A(A-2)}$ is the general HO bracket for two particles
with mass ratio equal to $A(A-2)$. The expression (\ref{taam1}) reduces to
Eq. (19) in Ref. \cite{NB99} for $A=4$. Let us stress the important
property of the antisymmetrizer matrix, namely its diagonality
in $N_{\rm A}\equiv N_{\rm L}=N_{\rm R}$. Consequently, we may
impose a model space restriction of the type $N_{\rm A}\leq N_{\rm max}$
and still obtain all the antisymmetrized states within that
model space.

The antisymmetrized states are obtained by diagonalizing the
antisymmetrizer ${\cal X}$ (\ref{xam11}) in the basis (\ref{abasam11})
or, more efficiently, by employing the method introduced in 
Ref. \cite{DK95}
that does not require us to compute all the matrix elements of the
antisymmetrizer. In order to get the basis for the $A$-nucleon system we
need to set up an iterative procedure that starts with the calculation
of the three-nucleon basis (\ref{abas3}) and procceeds to four nucleons
and so on up to $A$.
The resulting antisymmetrized states can be classified 
and expanded in terms of the original basis (\ref{abasam11}) 
similarly as in the three-nucleon case
\begin{eqnarray}\label{abasA}
|N_{\rm A} i_{\rm A} J T\rangle &=& \sum 
\langle N_{\rm A-1}i_{\rm A-1}J_{\rm A-1}T_{\rm A-1}; 
n_{\rm A-1}l_{\rm A-1}{\cal J}_{\rm A-1}||N_{\rm A} i_{\rm A} 
J T\rangle
\nonumber \\
&& 
|(N_{A-1} i_{A-1} J_{A-1} T_{A-1}; n_{A-1} l_{A-1} {\cal J}_{A-1}) 
JT\rangle \; ,
\end{eqnarray}
where $N_{\rm A}=N_{\rm A-1}+2n_{\rm A-1}+l_{\rm A-1}$. In Eq. (\ref{abasA}), 
the additional quantum number $i_{\rm A}$ distinguishes states with
the same set of quantum numbers $N_{\rm A}, J, T$, e.g.,
$i_A=1,2, \ldots r$ with 
$r$ the total number of antisymmetrized states for given 
$N_{\rm A}, J, T$.
Again, it can be obtained from computing the trace of the antisymmetrizer
${\cal X}$ 
\begin{equation}\label{traceA}
r= {\rm Tr}{\cal X}^{N_{\rm A} JT} \; .
\end{equation}

\subsection{Recoupling to basis containing two- and three-body clusters}
\label{recoupl}

The $A$-nucleon antisymmetrized basis, obtained as described in the previous
subsection, is given as an expansion of the basis (\ref{abasam11}), 
as shown in Eq. (\ref{abasA}). In this form, it is not, however, in general
suitable for calculations with two-body or three-body interactions.
In order to facilitate the calculations with two-body or three-body
interactions we need to expand the antisymmetrized states
$|N_{\rm A} i_{\rm A} JT\rangle$ in a HO basis consisting of
antisymmetrized subclusters of $A-2$ and 2 nucleons depending on the Jacobi
coordinates (\ref{jacobiam22}) or consisting of
antisymmetrized subclusters of $A-3$ and 3 nucleons depending on the
Jacobi coordinates (\ref{jacobiam33}), respectively.  

For a calculation for an $A>3$ system with a two-body interaction, 
we need the following expansion matrix element 
\begin{eqnarray}\label{recam22}
&&\langle (N_{\rm A-2} i_{\rm A-2} J_{\rm A-2} T_{\rm A-2}; (nlsjt,
{\cal N} {\cal L}){\cal J}) JT | 
N_{\rm A} i_{\rm A} JT \rangle \nonumber \\
&&=\sum
\langle N_{\rm A-1} i_{\rm A-1} J_{\rm A-1} T_{\rm A-1};\;
n_{\rm A-1} l_{\rm A-1}{\cal J}_{\rm A-1}||
N_{\rm A} i_{\rm A} JT\rangle \;
\nonumber \\
&&\times
\langle N_{\rm A-2} i_{\rm A-2} J_{\rm A-2} T_{\rm A-2};\;
n_{\rm A-2} l_{\rm A-2}{\cal J}_{\rm A-2}||
N_{\rm A-1} i_{\rm A-1} J_{\rm A-1} T_{\rm A-1}\rangle
\nonumber \\
&&\times 
\hat{\cal J}_{\rm A-1} 
\hat{\cal J}_{\rm A-2} \hat{\cal J} 
\hat{J}_{\rm A-1}\hat{j}\hat{s}
(-1)^{{\cal J}_{\rm A-2}+{\cal J}_{\rm A-1}+J_{\rm A-2}+J+j
+{\cal L}+s+l_{\rm A-1}+l_{\rm A-2}}
\nonumber \\
&&\times
(-1)^{T_{\rm A-2}+T+1}\hat{T}_{\rm A-1}\hat{t}
\left\{ \begin{array}{ccc} T_{\rm A-2} &\textstyle{\frac{1}{2}}  
               & T_{\rm A-1} \\
             \textstyle{\frac{1}{2}}  &  T  & t
\end{array}\right\}
\left\{ \begin{array}{ccc} l_{\rm A-2} & \textstyle{\frac{1}{2}} 
& {\cal J}_{\rm A-2}  \\
l_{\rm A-1} & \textstyle{\frac{1}{2}} 
& {\cal J}_{\rm A-1}  \\
L   & s  & {\cal J}
\end{array}\right\}
\left\{ \begin{array}{ccc} J_{\rm A-2}&{\cal J}_{\rm A-2}& J_{\rm A-1}\\
  {\cal J}_{\rm A-1}  & J & {\cal J}   
\end{array}\right\}
\left\{ \begin{array}{ccc} s   & l   & j   \\
  {\cal L}  & {\cal J}& L   
\end{array}\right\}
\nonumber \\
&&\times 
\hat{L}^2 \; 
\langle n l  {\cal N} {\cal L} \;L 
| n_{\rm A-1} l_{\rm A-1} n_{\rm A-2} l_{\rm A-2}  \;L
\rangle_{\frac{\rm A}{\rm A-2}}  \;  ,
\end{eqnarray}
where we used an orthogonal transformation of the Jacobi coordinates
$\vec{\xi}_{A-2},\vec{\xi}_{A-1}$ and $\vec{\eta}_{A-2},\vec{\eta}_{A-1}$.
In the state
$|(N_{\rm A-2} i_{\rm A-2} J_{\rm A-2} T_{\rm A-2}; (nlsjt,
{\cal N} {\cal L}){\cal J}) JT\rangle$, the antisymmetrized
subcluster $|N_{\rm A-2} i_{\rm A-2} J_{\rm A-2} T_{\rm A-2}\rangle$
depends on the Jacobi coordinates 
$\vec{\xi}_1,\vec{\xi}_{2},\ldots \vec{\xi}_{A-3}$. The two-nucleon
channel state $|nlsjt\rangle$ depends on the Jacobi coordinate 
$\vec{\eta}_{A-1}$ and the HO state $|{\cal NL}\rangle$ that describes
the relative motion of the two subclusters is associated with the
Jacobi coordinate $\vec{\eta}_{A-2}$, given in Eq. (\ref{jacobiam22}). 

When this expansion of $|N_{\rm A} i_{\rm A} JT\rangle$
is used a matrix element of a two-body interaction in the basis
$|N_{\rm A} i_{\rm A} JT\rangle$
can be evaluated in a simple manner, e.g.,
\begin{equation}\label{v2bme}
\langle \sum_{i<j=1}^A V_{ij} \rangle = \frac{1}{2}A(A-1)\langle
V(\sqrt{2}\vec{\eta}_{A-1})\rangle \; ,
\end{equation}
and the matrix element on the right-hand side is diagonal in all
quantum numbers of the state 
$|(N_{\rm A-2} i_{\rm A-2} J_{\rm A-2} T_{\rm A-2}; (nlsjt,
{\cal N} {\cal L}){\cal J}) JT\rangle$ except $n,l$ for an isospin 
invariant interaction.

Similarly, for a calculation for an $A>5$ system with 
a three-body interaction, we need the following expansion 
matrix element 
\begin{eqnarray}\label{recam33}
&&\langle (N_{\rm A-3} i_{\rm A-3} J_{\rm A-3} T_{\rm A-3}; 
(N_3 i_3 J_3 T_3,
{\cal N} {\cal L}){\cal J}) JT | 
N_{\rm A} i_{\rm A} JT \rangle \nonumber \\
&&=\sum
\langle N_{\rm A-2} i_{\rm A-2} J_{\rm A-2} T_{\rm A-2};\;
(nlsjt, \; {\cal N}' {\cal L}') {\cal J}'||
N_{\rm A} i_{\rm A} JT\rangle \;
\nonumber \\
&&\times
\langle N_{\rm A-3} i_{\rm A-3} J_{\rm A-3} T_{\rm A-3};\;
n_{\rm A-3} l_{\rm A-3}{\cal J}_{\rm A-3}||
N_{\rm A-2} i_{\rm A-2} J_{\rm A-2} T_{\rm A-2}\rangle \;
\nonumber \\
&&\times 
\langle nlsjt;\;
n'_{\rm A-3} l'_{\rm A-3}{\cal J}'_{\rm A-3}||
N_{\rm 3} i_{\rm 3} J_{\rm 3} T_{\rm 3}\rangle
\nonumber \\
&&\times 
\hat{\cal J}_{\rm A-3} 
\hat{{\cal J}'}_{\rm A-3} \hat{\cal J} \hat{{\cal J}'} 
\hat{J}_{\rm A-2}\hat{J}_3 \;
(-1)^{{\cal J}'+J_{\rm A-3}+J+j
+l_{\rm A-3}+\textstyle{\frac{1}{2}}}
\nonumber \\
&&\times
(-1)^{T_{\rm A-3}+T_3+T}\hat{T}_{\rm A-2}\hat{T}_3
\left\{ \begin{array}{ccc} T_{\rm A-3} &\textstyle{\frac{1}{2}}  
               & T_{\rm A-2} \\
            t  &  T  &T_3
\end{array}\right\}
\left\{ \begin{array}{ccc} J_{\rm A-3}   & J   & {\cal J}   \\
  {\cal J}' & {\cal J}_{\rm A-3} & J_{\rm A-2}
\end{array}\right\}
\nonumber \\
&&\times 
\hat{K}^2 (-1)^K
\left\{ \begin{array}{ccc} {\cal J}'_{\rm A-3} & J_3 & j  \\
l'_{\rm A-3} & {\cal L} & L  \\
\textstyle{\frac{1}{2}} & {\cal J} & K
\end{array}\right\}
\left\{ \begin{array}{ccc} {\cal J}'  & l_{\rm A-3} & K    \\
  \textstyle{\frac{1}{2}} & {\cal J}  & {\cal J}_{\rm A-3}
\end{array}\right\}
\left\{ \begin{array}{ccc} {\cal J}'   & l_{\rm A-3}   & K   \\
  L & j & {\cal L}'
\end{array}\right\}
\nonumber \\
&&\times 
\hat{L}^2 (-1)^L\; 
\langle n'_{\rm A-3} l'_{\rm A-3}  {\cal N} {\cal L} \;L 
| n_{\rm A-3} l_{\rm A-3} {\cal N}' {\cal L}'  \;L
\rangle_{\frac{\rm 2(A-3)}{\rm A}}  \; .
\end{eqnarray}
Here we used an orthogonal transformation of the Jacobi coordinates
$\vec{\xi}_{A-3},\vec{\eta}_{A-2}$ and 
$\vec{\vartheta}_{A-3},\vec{\vartheta}_{A-2}$ given in 
Eqs. (\ref{jacobiam22},\ref{jacobiam33}).
In the state
$|(N_{\rm A-3} i_{\rm A-3} J_{\rm A-3} T_{\rm A-3}; 
(N_3 i_3 J_3 T_3,{\cal N} {\cal L}){\cal J}) JT\rangle$, 
the antisymmetrized subcluster 
$|N_{\rm A-3} i_{\rm A-3} J_{\rm A-3} T_{\rm A-3}\rangle$
depends on the Jacobi coordinates 
$\vec{\xi}_1,\vec{\xi}_{2},\ldots \vec{\xi}_{A-4}$. The three-nucleon
antisymmetrized subcluster $|N_3 i_3 J_3 T_3\rangle$ 
depends on the Jacobi coordinates 
$\vec{\vartheta}_{A-2},\vec{\eta}_{A-1}$ 
and the HO state $|{\cal NL}\rangle$ that describes
the relative motion of the two subclusters is associated with the
Jacobi coordinate $\vec{\vartheta}_{A-3}$, given in Eq. (\ref{jacobiam33}). 

When this expansion of $|N_{\rm A} i_{\rm A} JT\rangle$
is used a matrix element of a three-body interaction in the basis
$|N_{\rm A} i_{\rm A} JT\rangle$
can be evaluated in a straightforward way, e.g.,
\begin{equation}\label{v3bme}
\langle \sum_{i<j<k=1}^A V_{ijk} \rangle = \frac{1}{6}A(A-1)(A-2)\langle
V(\vec{\vartheta}_{A-2},\vec{\eta}_{A-1})\rangle \; ,
\end{equation}
and the matrix element on the right-hand side is diagonal in all
quantum numbers of the state 
$|(N_{\rm A-3} i_{\rm A-3} J_{\rm A-3} T_{\rm A-3}; 
(N_3 i_3 J_3 T_3,{\cal N} {\cal L}){\cal J}) JT\rangle$, 
except $N_3$ and $i_3$, for an isospin invariant interaction.

\subsection{One-body densities}
\label{densi}

In the formalism of the translationally-invariant shell model,
one can introduce one- or higher-body densities, 
as in the standard shell-model formulation. Let us consider
a general one-body operator, e.g.,
\begin{equation}\label{obop}
\hat{O}^{(k\tau)}=\sum_{i=1}^A \hat{O}^{(k\tau)}(\vec{r}_i-\vec{R},
\vec{\sigma}_i,\vec{\tau}_i)  \; .
\end{equation}
Its matrix element between antisymmetrized states depending
on the Jacobi coordinates (\ref{jacobiam11}) can be written 
schematically as
\begin{equation}\label{obopme}
\langle \hat{O}^{(k\tau)} \rangle = A \langle \hat{O}^{(k\tau)}
(-\sqrt{\frac{A-1}{A}}\vec{\xi}_{A-1},\vec{\sigma}_A,\vec{\tau}_A)
\rangle
\; .
\end{equation}
In a more detailed form, we can express
a reduced matrix element between two eigenstates of a Hamiltonain
corresponding to $A$-nucleon system as 
\begin{eqnarray}\label{obdme}
&&\langle A; E J^\pi T |||\hat{O}^{(k\tau)}||| A; E' J'^{\pi'} T'
\rangle 
\nonumber \\
&&=\sum
\langle A; E J^\pi T |(N_{\rm A-1} i_{\rm A-1} J_{\rm A-1} T_{\rm A-1};
n_{\rm A-1} l_{\rm A-1} {\cal J}_{\rm A-1}) JT\rangle \;
\nonumber \\
&&\times
\langle (N_{\rm A-1} i_{\rm A-1} J_{\rm A-1} T_{\rm A-1};
n'_{\rm A-1} l'_{\rm A-1} {\cal J}'_{\rm A-1}) J' T'|
A; E' J'^{\pi'}T'\rangle
\nonumber \\
&&\times
\hat{J}\hat{J}' (-1)^{J_{\rm A-1}+K+J+{\cal J}'_{\rm A-1}}
\left\{ \begin{array}{ccc} J_{\rm A-1} & {\cal J}'_{\rm A-1} & J' \\
  K  & J & {\cal J}_{\rm A-1}   
\end{array}\right\}
\hat{T}\hat{T}' (-1)^{T_{\rm A-1}+\tau+T+\textstyle{\frac{1}{2}}}
\left\{ \begin{array}{ccc} T_{\rm A-1} &\textstyle{\frac{1}{2}}  
               & T' \\
             \tau  &  T  & \textstyle{\frac{1}{2}}  
\end{array}\right\}
\nonumber \\
&&\times
\langle n_{\rm A-1} l_{\rm A-1} {\cal J}_{\rm A-1} |||
A \hat{O}^{k\tau}
(-\sqrt{\frac{A-1}{A}}\vec{\xi}_{A-1},\vec{\sigma}_A,\vec{\tau}_A)
|||n'_{\rm A-1} l'_{\rm A-1} {\cal J}'_{\rm A-1}
\rangle
\; ,
\end{eqnarray}
where we used expansions of the eigenstates in the basis (\ref{abasA}).
Apparently, the matrix element (\ref{obdme}) factorizes into products
of one-body reduced matrix elements and one-body densities.

One can introduce two- or higher-body densities in a similar way.

\section{Results for few-nucleon systems}
\label{sec4}

We have written a computer code for calculations using the formalism
presented in Sections \ref{sec2} and \ref{sec3}. We performed test 
calculations for few-nucleon systems up to $A=8$. The code reproduces
results obtained using the many-fermion dynamics (MFD)
shell-model code \cite{VZ94},
when a two-body effective interaction is employed and when the model space
does not require more than 9 major HO shells, i.e., the limits
of the version of the MFD code we have. Due to the computational complexity
of the antisymmetrization procedure we expect that the present formalism
can, at the present time and in the current formulation, improve significantly 
on the results obtainable by the MFD code for $A\leq 6$ and to some extent
for $A=7$ and $A=8$. Let us further remark that although the antisymmetrization 
is complicated it needs to be done only once for given $A$, $N_{A}$, $J$ 
and $T$.

In this paper we present results for $A=3$ and $A=4$ systems,
while calculations for larger $A$ will be published separately
at a later stage. We investigated the $A=3$ and $A=4$ systems in the
framework of the present formalism in two previous papers \cite{NB98} and 
\cite{NB99}, respectively. However, as the newly developed code is more
efficient, we were able to extend the calculations to larger model spaces.
In addition, in this paper we present results obtained using 
the momentum-space dependent non-local CD-Bonn NN potential
\cite{Machl}. There are no published results for the four-nucleon system
interacting by CD-Bonn potential up to now. 

We work in the isospin formalism. As the CD-Bonn NN potential breaks
the isospin and charge symmetry we construct an isospin invariant potential 
for the $T=1$ two-nucleon channels by taking
combinations of $V_{\rm np}$, $V_{\rm pp}$ and $V_{\rm nn}$. For $^3$H
we take $\frac{1}{3}V_{\rm np}+\frac{2}{3}V_{\rm nn}$, for $^3$He we
take $\frac{1}{3}V_{\rm np}+\frac{2}{3}V_{\rm pp}$, while for $^4$He
we use $\frac{1}{3}V_{\rm np}+\frac{1}{3}V_{\rm pp}+\frac{1}{3}V_{\rm nn}$.
The Coulomb potential is added to the CD-Bonn $V_{\rm pp}$ potential.
Similarly, for the nucleon mass we use the proton and neutron
mass combinations, e.g.,
$m=\frac{1}{A}(Z m_{\rm p} + N m_{\rm n})$ with $Z$ and $N$ the number
of protons and neutrons, respectively.

\subsection{$^3$H and $^3$He} 

For the $A=3$ system we use the two-body effective
interaction calculated as described in Subsection \ref{v2beff}.
As the effective interaction depends on the model-space size,
characterized by $N_{\rm max}$, and
the HO frequency $\Omega$, we investigate the dependence of observables
on those two parameters.
We performed calculations in the model spaces
with $N_{\rm max}$ up to 34 for a wide range of HO frequencies $\Omega$,
e.g., $\hbar\Omega=19 \ldots 32$ MeV. 
Our ground-state results are presented in Figs. \ref{fig3Hf}-\ref{fig3He}. 
In Fig. \ref{fig3Hf} we show the $^3$H ground-state energy dependence
on the model-space size in the range of $N_{\rm max}=4 - 34$. Different
full lines connect results obtained with different HO frequencies.
The dotted line represents the 34-channel Faddeev equation result
-8.00 MeV \cite{Machl}. It is apparent that our results converge  
to the Faddeev equation result as $N_{\rm max}$ increases. We note
that the fundamental approximation used in our approach is the negligence
of the three-body clusters in the expansion (\ref{UMOAexpan}).
Such clusters can give both positive and negative contribution to the
ground-state energy. 
Our calculation is not a variational calculation. Therefore, we cannot
expect a convergence from above. As seen from Fig. \ref{fig3Hf} our results
converge both from above or below, with some oscillations possible,
depending on the HO frequency employed. In Fig. \ref{fig3H} we present 
the same as in Fig. \ref{fig3Hf} for $N_{\rm max}$ in the range
from 14 to 34 using an expanded energy scale. We can see that, even if
a complete convergence has not been achieved for $N_{\rm max}=34$
in the whole range of $\Omega$ used, it is possible to interpolate
from the different curves to obtain the converged result. For example,
the line corresponding to $\hbar\Omega=28$ MeV remains almost constant
for $N_{\rm max}>18$ at the value about -8.00 MeV. We present the 
interpolated results in Table \ref{tab1}. Our $^3$H result -8.002(4) MeV
is in a good agreement with the 34-channel Faddeev calculation. We note,
however, that in our calculations we used all the two-nucleon channels
up to $j=6$. We should, therefore, compare with the result -8.014 MeV
obtained by Nogga et al. \cite{NHKG97}, where all channels with $j\leq 6$ were
used. Consequently, it appears that we are missing about 10 keV in binding,
most likely due to imprecisions in the two-nucleon systems solutions used
for construction of the effective interaction.

In Fig. \ref{fig3He}, we present the same dependence for $^3$He as 
in Fig. \ref{fig3H}
for $^3$H. Our interpolated ground-state energy result together 
with the error estimate is given in Table \ref{tab1}. It should be noted
that the CD-Bonn potential gives a realistic prediction for the
binding energy difference of $^3$H and $^3$He, which is experimentaly 0.764 MeV. 
On the other hand, the absolute
value of the binding energy is underestimated compared to the experimental
values, 8.482 MeV for $^3$H and 7.718 MeV for $^3$He, by about 400 keV. 
It is, however, only a half of what one gets
with the local coordinate-space potentials like Nijmegen, Reid or Argonne.

The model-space size and $\Omega$ dependence of the $^3$H point-nucleon
matter rms radius is presented in Fig. \ref{fig3Hradf}. We observe a 
convergence and a saturation with the model-space size increase.
In Table \ref{tab1} we show point-proton and point-neutron rms radii
for both $^3$H and $^3$He extrapolated from the calculations in the largest
model spaces together with the error estimates. In addition, the
calculated magnetic moments are also presented. These can be compared to the
experimental values of $+2.979\; \mu_{\rm N}$ for $^3$H and $-2.128 \;\mu_{\rm N}$
for $^3$He. The calculated values were obtained using a one-body M1 operator
with bare nucleon $g$-factors.

\subsection{$^4$He} 

The calculations for $^4$He were performed in the model spaces
up to $N_{\rm max}=16$ in a wide range of HO frequencies $\Omega$.
This is an extension of our previous $A=4$ calculations \cite{NB99},
where model spaces only up to $N_{\rm max}=14$ were utilized and
a narrower range of $\Omega$ was investigated.
We performed separate calculations both with 
two-body effective interactions, computed as described 
in Subsection \ref{v2beff}, and with three-body effective interactions,
computed as discussed in Subsection \ref{v3beff}. The three-body
effective interactions were calculated for the three-nucleon channels
with $J_3^\pi T_3=\frac{1}{2}^\pm \frac{1}{2}$ and  
$J_3^\pi T_3=\frac{3}{2}^\pm \frac{1}{2}$ using $N_{\rm 3max}=32$ and
$N_{\rm 3max}=28$, respectively. 

Our ground-state energy results are
presented in Figs. \ref{fig4He} and \ref{fig4HeOmega}. 
We investigate the dependence on both $N_{\rm max}$ and $\Omega$.
The two figures show mostly the same points, plotted in the first
case as a function of $N_{\rm max}$ and in the second case as a function
of $\hbar\Omega$. The dotted lines
connect the results obtained using the two-body effective interactions,
while the full lines connect the results obtained with the three-body
effective interactions. We observe a decrease of dependence on both
$N_{\rm max}$ and $\hbar\Omega$ as the model-space size increases.
In particular, the curves in Fig. \ref{fig4HeOmega} become more flat
with increasing $N_{\rm max}$. Similarly as in our previous study \cite{NB99},
it is apparent that calculations done with the three-body effective interaction
show weaker dependence on both $\Omega$ and $N_{\rm max}$ and
demonstrate faster convergence. Due to the higher complexity 
of those calculations,
we present results only for $\hbar\Omega=13,22,31$ and 40 MeV.
As can be seen from Figs. \ref{fig4He} and \ref{fig4HeOmega},
for $\hbar\Omega$ less than about 40 MeV the binding energy
decreases with increasing model-space size while for larger
$\hbar\Omega$ it begins to increase with increasing $N_{\rm max}$. 
Similarly as for the $A=3$ system, we are in a position to interpolate
the converged ground-state energy result, though with a lower accuracy.
Based on the results presented in Figs. \ref{fig4He} and \ref{fig4HeOmega},
we estimate the CD-Bonn $^4$He ground-state energy to be -26.4(2) MeV.
The experimental binding energy of $^4$He is 28.296 MeV. The CD-Bonn thus
underbinds $^4$He by about 2 MeV. It is again only about a half
of underbinding that one gets with, e.g., Argonne V18 with the calculated
$^4$He binding energy 24.1 MeV \cite{CS98}.

In our approach we obtain the ground state
as well as the excited states by diagonalizing the Hamiltonian.
In Figs. \ref{fig4Heexc} and \ref{fig4Herad}, 
we present the model-space-size dependence
of both the ground-state and the first excited $0^+ 0$ state energies
and point-nucleon rms radii, respectively,
obtained in calculations with the three-body effective interactions.
Compared to the ground state, we observe
a much stronger dependence by the excited-state energy and 
nucleon rms radius
on both $\Omega$ and $N_{\rm max}$. 
The significantly different convergence rate of the ground state
and of the first excited $0^+ 0$ state manifests the different 
nature of the two states.
A possible interpretation of this observation is
that the excited $0^+ 0$ state is associated with a radial excitation
and, thus, it is more sensitive to the HO basis used in our calculations.
Although we cannot extrapolate the energy or point-nucleon radius 
of this state,
our calculations show that its excitation energy is below 21 MeV 
and the radius is larger than 3 fm.

Our ground-state results are summarized in Table \ref{tab1}.

\subsection{Charge form factors}

A sensitive test of the wave-functions obtained in our
calculations is the evaluation of form factors. 
In this subsection we compare the charge form factors obtained
with the CD-Bonn wave functions and those obtained in an identical
calculation with the Argonne V8' NN potential defined in 
Ref. \cite{GFMC}. 
We note that a similar comparison was performed by Kim et al.
in Ref. \cite{K88} for the Bonn OBEPQ and the Reid NN potentials.
In that paper, it was found that the $^3$H and $^3$He form factors
differ for the two NN potentials.  

Using the formalism
of Ref. \cite{MSD94}, we calculated the charge EM form factors
and ratio of charge strangeness and EM
form factors in the impulse approximation. The one-body contribution
to the charge operator is given by Eq. (15) in Ref. \cite{MSD94}, e.g.,
\begin{equation}\label{chargeop}
\hat{M}_{00}^{(a)}(q)^{[1]}=\frac{1}{2\sqrt{\pi}}\sum_{k=1}^A
\left\{\frac{G_E^{(a)}(\tau)}{\sqrt{1+\tau}}j_0(q r_k)+
\left[ G_E^{(a)}(\tau) - 2 G_M^{(a)}(\tau) \right] 
2\tau \frac{j_1(q r_k)}{q r_k} {\bf \sigma}_k \cdot {\bf L}_k
\right\} \; ,
\end{equation}
where $\tau=\frac{q^2}{4 m_N ^2}$, ${\bf L}_k$ is the $k$-th nucleon
orbital momentum, $G_E^{(a)}(\tau)$ and $G_M^{(a)}(\tau)$ 
are the one-body electric and magnetic form factors, respectively. 
The superscript $(a)$ refers to $(p)$ and $(n)$ for proton and neutron 
EM form factor, respectively, 
or to $(s)$ for the strangeness form factor. We use 
the parametrization of the one-body form factors as 
discussed in Ref. \cite{MSD94}.
We note that the one-body strangeness form factors depend on
the strangeness radius $\rho_s$, for which we take the value $\rho_s =-2.0$ 
as in Ref. \cite{MSD94} and on the strangeness magnetic moment $\mu_s$.
Limits on these parameters are to be determined in the experiments
at the Thomas Jefferson Accelerator Facility (TJNAF). 
The first strangeness magnetic-moment measurement
was reported recently \cite{M97} and an experimental value 
$\mu_s=+0.23$,
obtained with a large error. We use this value in our calculations.

The elastic EM charge form factors of $^3$H and $^3$He are presented
in Figs. \ref{figformf3H} and \ref{figformf3He}, respectively.
We observe a large sensitivity to the choice of the NN potential.
Let us remark that we used the wave functions obtained in the model
space with $N_{\rm max}=34$. We investigated the dependence of the
form factors on both $N_{\rm max}(=30,32,34)$ and $\hbar\Omega$ and 
found that the dependence is below the resolution of the figures.
Our Argonne V8' results compare well with those obtained using the Argonne V18
in the impulse approximation presented, e.g., in Ref. \cite{CS98}. 
We note that the experimental position of the minima as about 3.6 fm$^{-1}$
and 3.2 fm$^{-1}$ for $^3$H and $^3$He, respectively, see Ref. \cite{CS98}
and the references therein.

The $^4$He elastic EM charge form factor and the EM charge form factor 
corresponding to the transition to the first excited $0^+ 0$ state
are presented in Figs. \ref{figformf4He} and \ref{figformf4Hetr},
respectively. The three-body effective interactions were used
and the $N_{\rm max}=16$ model spaces. For the Argonne V8', the
current results can be compared to those presented in Ref. \cite{NB99}
obtained using $N_{\rm max}=14$. 
We note that a second minimum appears in our calculated
charge form factors in a similar position as found in the
VMC calculations presented in Ref. \cite{SPR90}.   
The elastic charge form factor sensitivity to both $N_{\rm max}$
and $\hbar\Omega$ is weak for $q$ below the secondary maximum
but it increases for larger $q$. As to the inelastic form factor, there
the sensitivity is significantly stronger. We believe that the 
form factors presented in Fig. \ref{figformf4Hetr} are more realistic
than the inelastic ones given in Ref. \cite{NB99} that we obtained using
$N_{\rm max}=14$.

In general, for all $^3$H, $^3$He and $^4$He the CD-Bonn results 
are further from the experimental data points than the results
obtained using the Argonne V8'. This is in full
agreement with the calculations in Ref. \cite{K88} for $^3$H and 
$^3$He. 
However, in order to make any conclusion about the superiority of any
of the potentials, one needs to calculate the meson exchange current
contributions.

Finally, in Fig. \ref{figformf4Her} we present the ratio of the $^4$He
charge strangeness and EM form factors calculated in the impulse 
approximation using both the Argonne V8' and the CD-Bonn NN potentials.
The ratio of the elastic charge form factors is particularly interesting,
as it can be experimentaly obtained from the measurement of the
parity-violating left-right asymmetry for scattering of polarized
electrons from a $^4$He target. Experiments of this type are now under
preparation at TJNAF.

\section{Conclusions}
\label{sec5}

We presented a translationally invariant formulation of the no-core 
shell-model approach for few-nucleon systems and
introduced a general method of antisymmetrization
of a HO basis depending on Jacobi coordinates.  
The latter procedure starts with the construction of the antisymmetrized basis
for three nucleons, then procceeds to four and so on.  
We derived an iterative algebraic formula for computing the antisymmetrized
basis for $A$ nucleons from the antisymmetrized basis for $A-1$ nucleons.
In addition, we discussed how to transform the antisymmetrized states
to bases containing different antisymmetrized subclusters of nucleons.
The chosen approach has the advantage that the antisymmetrizer is very
simple and that the dimensions of the starting basis, formed
by the $A-1$ nucleon antisymmetrized subcluster and the last nucleon, 
are the lowest compared to bases with different subclustering. 

There are two main advantages of the use of a translationally-invariant
basis. First, it allows us to employ larger 
model spaces than in traditional shell-model calculations. 
Second, in addition to two-body effective interactions, 
three- or higher-body effective interactions as well as real three-body 
interactions can be utilized. The use of higher-body effective interactions
reduces the dependence on the HO frequency and speeds up the convergence 
of our approach.

As the antisymmetrization
procedure is computationally involved, the practical applicability of the
formalism is limited to light nuclei. In the present formulation we expect
that significant improvement over the traditional shell-model results
can be achieved for $A\leq 6$ and to some extent for $A=7$ and $A=8$.
However, different paths of antisymmetrization
than that realized here can be chosen for more complex nuclei.
On the other hand, there exists a sophisticated approach for
antisymmetrization of hyperspherical functions depending on Jacobi 
coordinates developed by Barnea and Novoselsky \cite{BN97}. It makes use
of the orthogonal group transformations of Jacobi coordinates. That approach 
can be adapted for the HO functions and should lead to a more efficient
antisymmetrization. Another issue is the transformation to $A-2$ plus 2 and
$A-3$ plus 3 clusters as discussed in Subsection \ref{recoupl}. In principle,
it can be avoided by using the reverse ordering of the Jacobi coordinates
(\ref{jacobiam11}) \cite{B99}. Then, however, the computation of the
one-body densities, as described in Subsection \ref{densi}, would become very
difficult.

We applied the formalism to solve three and four nucleon systems 
interacting by the CD-Bonn NN potential in model spaces that included 
up to $34\hbar\Omega$ and $16\hbar\Omega$ HO excitations, respectively.
For the three-nucleon system our method leads to the exact solution
and our results are in agreement with the calculations by other methods.
For the four-nucleon system, we were able to interpolate the ground-state
energy solution from the model-space and HO-frequency dependencies.
Our result with the error estimate is -26.4(2) MeV. 
There have not been any published results by other methods so far for 
the $A=4$ system interacting by the CD-Bonn NN potential to which we could
compare.
In addition to energies, rms radii and magnetic moments, we also
compared charge form factors obtained using the CD-Bonn and 
Argonne V8' NN potentials. We found a substantial sensitivity to 
the choice of
the potential in agreement with results published in Ref. \cite{K88} 
calculated with similar potentials.

We believe that the method discussed in this paper has the potential
to solve the few-nucleon problem beyond $A=4$. The convergence can still
be improved by employing four- or higher-body effective interactions
in a similar fashion as we used the three-body effective interaction.
Also the antisymmetrization procedure can be made more efficient.
Our method has the advantage, compared to, e.g., the GFMC method,
that we solve the Schr\"odinger equation by diagonalization. Consequently,
wave functions are obtained and excited states with identical quantum 
numbers as the ground state are computed. 
It would be also interesting to compare the present method with
a new approach, relying also on the HO basis, in which the effective
interaction is constructed by solving the Bloch-Horowitz equation \cite{HS99}. 
Calculations are now under way for $A=5$ and $A=6$ using the present 
formalism.

\acknowledgements{We would like to thank R. Machleidt for providing
the computer code for the CD-Bonn NN potential.
We also thank Kenji Suzuki and Nir Barnea for useful discussions.
This work was supported in part by the NSF grants No. PHY96-05192
and INT98-06614. G.P.K. acknowledges the CIES for a Fulbright
Research Fellowship while at the University of Arizona and partial
support from the grant No. 391 of the Lithuanian State Science
and Studies Foundation.
}

\begin{figure}
\caption{The dependence of the $^3$H ground-state energy, 
in MeV, on the maximal number of HO excitations allowed
in the model space in the range from $N_{\rm max}=4$ 
to $N_{\rm max}=34$. The two-body effective interaction 
utilized was derived from the CD-Bonn NN potential. 
Results for $\hbar\Omega=19, 22, 24, 26, 28, 30$ and 32 MeV 
are presented. The dotted line represents the exact result of 
-8.00 MeV from a 34-channel
Faddeev-equation calculation \protect\cite{Machl}.
}
\label{fig3Hf}
\end{figure}

\begin{figure}
\caption{The dependence of the $^3$H ground-state energy, 
in MeV, on the maximal number of HO excitations allowed
in the model space in the range from $N_{\rm max}=14$ 
to $N_{\rm max}=34$. The two-body effective interaction 
utilized was derived from the CD-Bonn NN potential. 
Results for $\hbar\Omega=19, 22, 24, 26, 28, 30$ and 32 MeV 
are presented. The dotted line represents the exact result of 
-8.00 MeV from a 34-channel
Faddeev-equation calculation \protect\cite{Machl}.
}
\label{fig3H}
\end{figure}

\begin{figure}
\caption{The dependence of the $^3$He ground-state energy, 
in MeV, on the maximal number of HO excitations allowed
in the model space in the range from $N_{\rm max}=14$ 
to $N_{\rm max}=34$. The two-body effective interaction 
utilized was derived from the CD-Bonn NN potential. 
Results for $\hbar\Omega=19, 22, 24, 26, 28$ and 30 MeV 
are presented.
}
\label{fig3He}
\end{figure}

\begin{figure}
\caption{The dependence of the $^3$H point-nucleon matter 
radius, in fm, on the maximal number of HO excitations allowed
in the model space in the range from $N_{\rm max}=4$ 
to $N_{\rm max}=34$. The two-body effective interaction 
utilized was derived from the CD-Bonn NN potential. 
Results for $\hbar\Omega=19, 22, 24, 26, 28, 30$ and 32 MeV 
are presented. 
}
\label{fig3Hradf}
\end{figure}

\begin{figure}
\caption{The dependence of the $^4$He ground-state energy, 
in MeV, on the maximal number of HO excitations allowed
in the model space in the range from $N_{\rm max}=8$ 
to $N_{\rm max}=16$. The two-body (dotted lines) and three-body
(full lines) effective interactions 
utilized were derived from the CD-Bonn NN potential. 
Results for $\hbar\Omega=19, 22, 31, 37,
40$ and 43 MeV are presented.
}
\label{fig4He}
\end{figure}

\begin{figure}
\caption{The dependence of the $^4$He ground-state energy, 
in MeV, on the HO energy in the range from $\hbar\Omega=13$ 
to $\hbar\Omega=43$. The two-body (dotted lines) and three-body
(full lines) effective interactions 
utilized were derived from the CD-Bonn NN potential. 
Results for $N_{\rm max}=8, 10, 12, 14$
and 16 are presented.
}
\label{fig4HeOmega}
\end{figure}

\begin{figure}
\caption{The dependence of the $^4$He ground-state and the 
first-excited $0^+ 0$ state energies, in MeV, 
on the maximal number of HO excitations allowed
in the model space in the range from $N_{\rm max}=8$ 
to $N_{\rm max}=16$. The three-body effective interaction 
utilized was derived from the CD-Bonn NN potential. 
Results for $\hbar\Omega=13, 22, 31$ and 40 MeV are presented.
}
\label{fig4Heexc}
\end{figure}

\begin{figure}
\caption{The dependence of the $^4$He ground-state and the 
first-excited $0^+ 0$ state point-nucleon rms radius, in fm, 
on the maximal number of HO excitations allowed
in the model space in the range from $N_{\rm max}=8$ 
to $N_{\rm max}=16$. The three-body effective interaction 
utilized was derived from the CD-Bonn NN potential. 
Results for $\hbar\Omega=13, 22, 31$ and 40 MeV are presented.
}
\label{fig4Herad}
\end{figure}

\begin{figure}
\caption{The elastic EM charge form factor of $^3$H calculated 
in the impulse approximation. Results obtained using the Argonne 
V8' (dotted line) and CD-Bonn (full line) NN potentials are
compared.
}
\label{figformf3H}
\end{figure}

\begin{figure}
\caption{The elastic EM charge form factor of $^3$He calculated 
in the impulse approximation. Results obtained using the Argonne 
V8' (dotted line) and CD-Bonn (full line) NN potentials are
compared.
}
\label{figformf3He}
\end{figure}

\begin{figure}
\caption{The elastic EM charge form factor of $^4$He calculated 
in the impulse approximation. Results obtained using the Argonne 
V8' (dotted line) and CD-Bonn (full line) NN potentials are
compared. The calculations were performed using three-body 
effective interaction in the model space
characterized by $N_{\rm max}=16$ and $\hbar\Omega=22$ MeV.
}
\label{figformf4He}
\end{figure}

\begin{figure}
\caption{The EM charge form factor of $^4$He corresponding to the
transition to the first excited $0^+ 0$ state calculated 
in the impulse approximation. Results obtained using the Argonne 
V8' (dotted line) and CD-Bonn (full line) NN potentials are
compared. The calculations were performed using three-body 
effective interaction in the model space
characterized by $N_{\rm max}=16$ and $\hbar\Omega=22$ MeV.
}
\label{figformf4Hetr}
\end{figure}

\begin{figure}
\caption{The ratio of elastic strangeness and EM charge form factor 
of $^4$He calculated 
in the impulse approximation. Results obtained using the Argonne 
V8' (dotted line) and CD-Bonn (full line) NN potentials are
compared. The calculations were performed using three-body 
effective interaction in the model space
characterized by $N_{\rm max}=16$ and $\hbar\Omega=22$ MeV.
Values of the strangeness radius $\rho_s=-2.0$ and 
the strangeness magnetic moment $\mu_s=0.23$ were employed.
}
\label{figformf4Her}
\end{figure}

\begin{table}
\begin{tabular}{ccccc}
CD-Bonn NN potential 
& $E_{\rm gs}$ [MeV]
& $\sqrt{\langle r^2_{\rm p} \rangle}$ [fm] 
& $\sqrt{\langle r^2_{\rm n} \rangle}$ [fm]
& $\mu$ [$\mu_{\rm N}$] \\ 
\hline
$^3$H & -8.002(4) & 1.608(4) & 1.760(6) & 2.612 \\
$^3$He & -7.248(4) & 1.802(6) & 1.635(4) & -1.779 \\
$^4$He & -26.4(2) & 1.445(5) & 1.445(5)  & -
\end{tabular}
\caption{Results for the ground-state energies, point-proton 
and point-neutron rms radii, and magnetic moments  
obtained for $^3$H, $^3$He and $^4$He using the CD-Bonn NN potential
are presented.
Shown values are based on the results calculated in the largest model spaces 
used in the present study, $N_{\rm max}=34$ for $^3$H, $^3$He, and 
$N_{\rm max}=16$ for $^4$He, respectively. The errors were estimated
from the dependences on the HO frequency $\Omega$ and on the model-space 
size characterized by $N_{\rm max}$.  
}
\label{tab1}
\end{table}

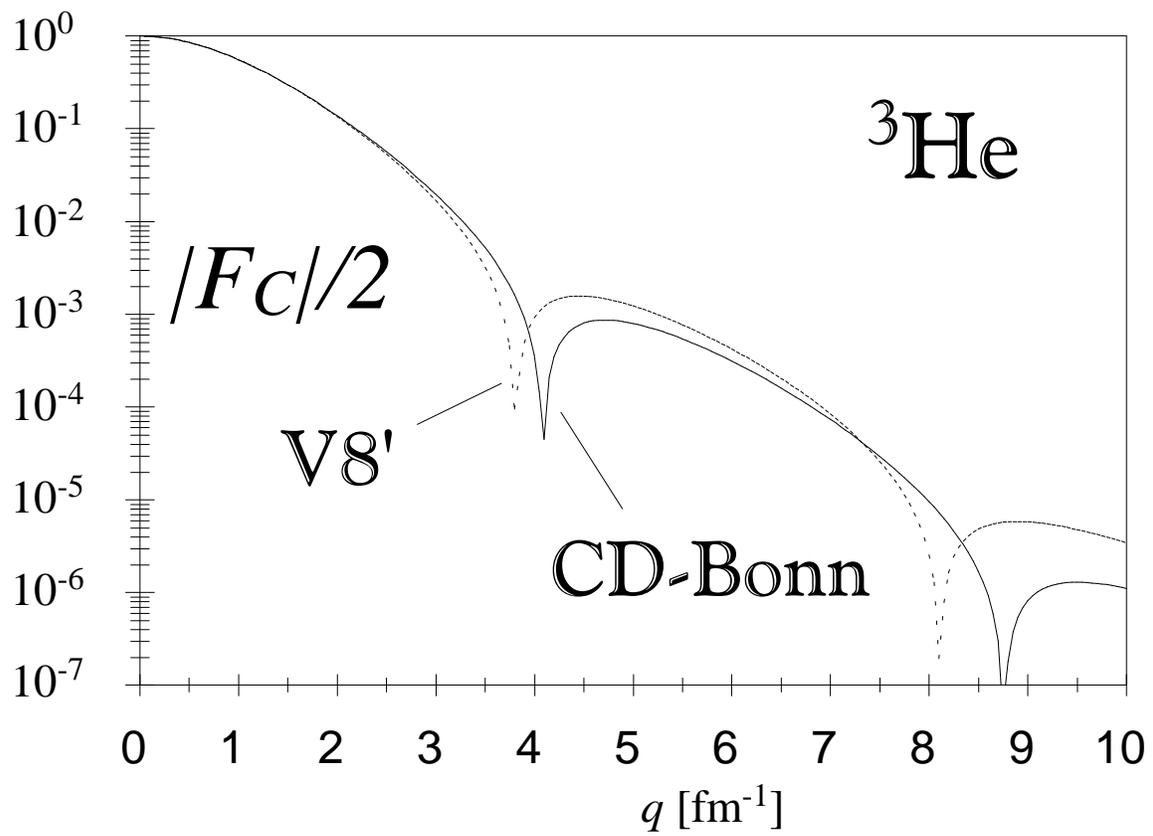
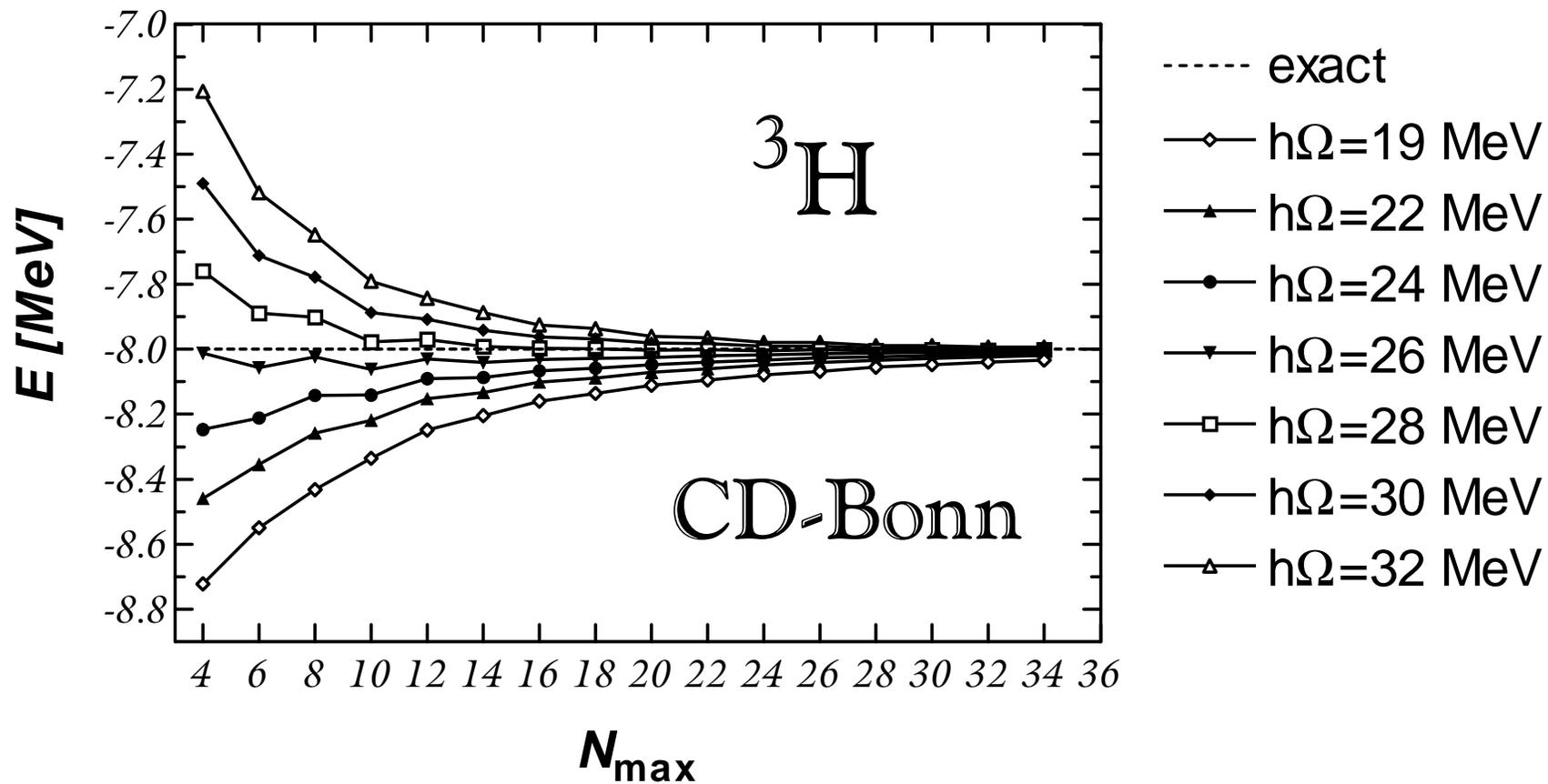
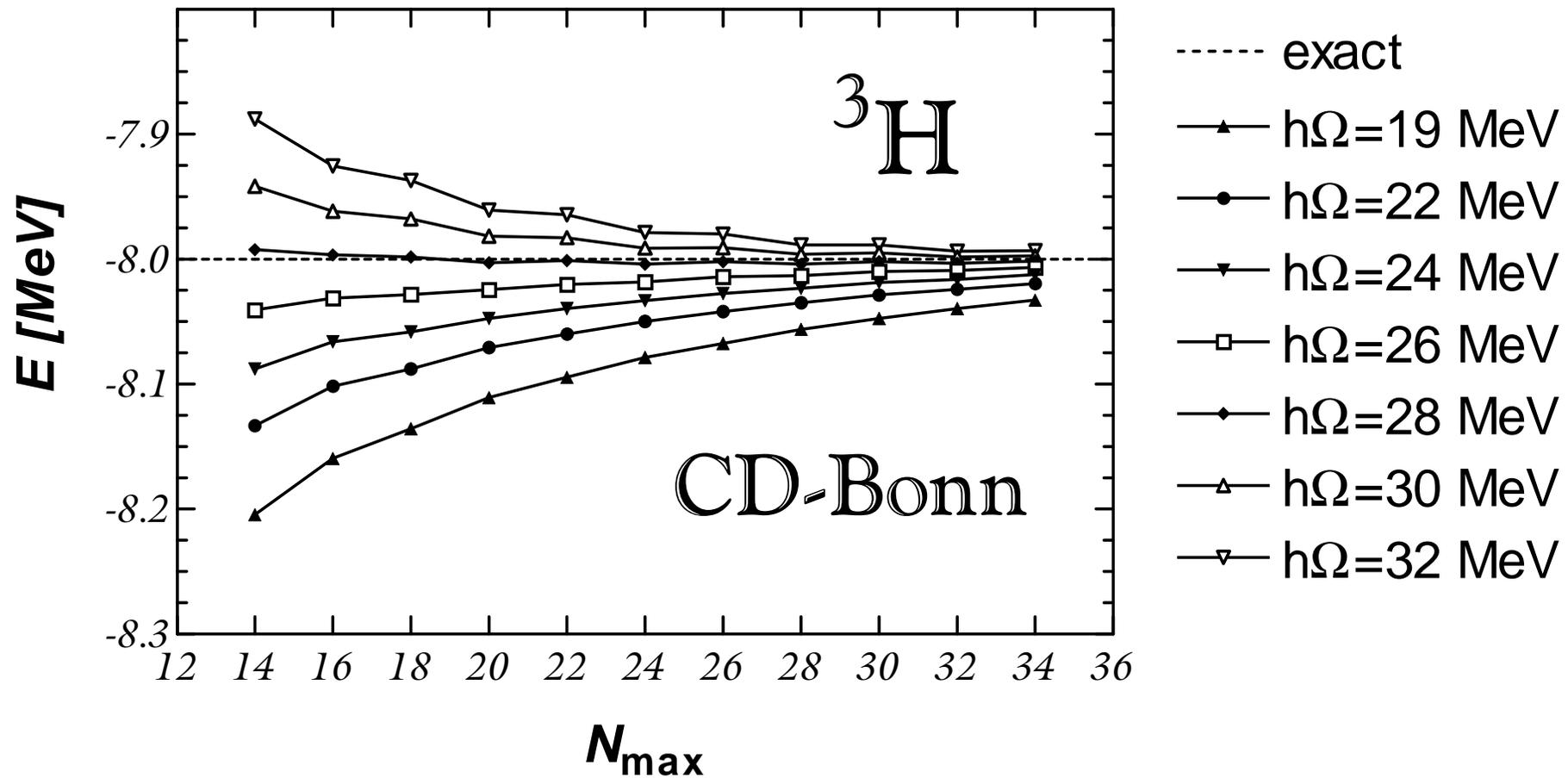
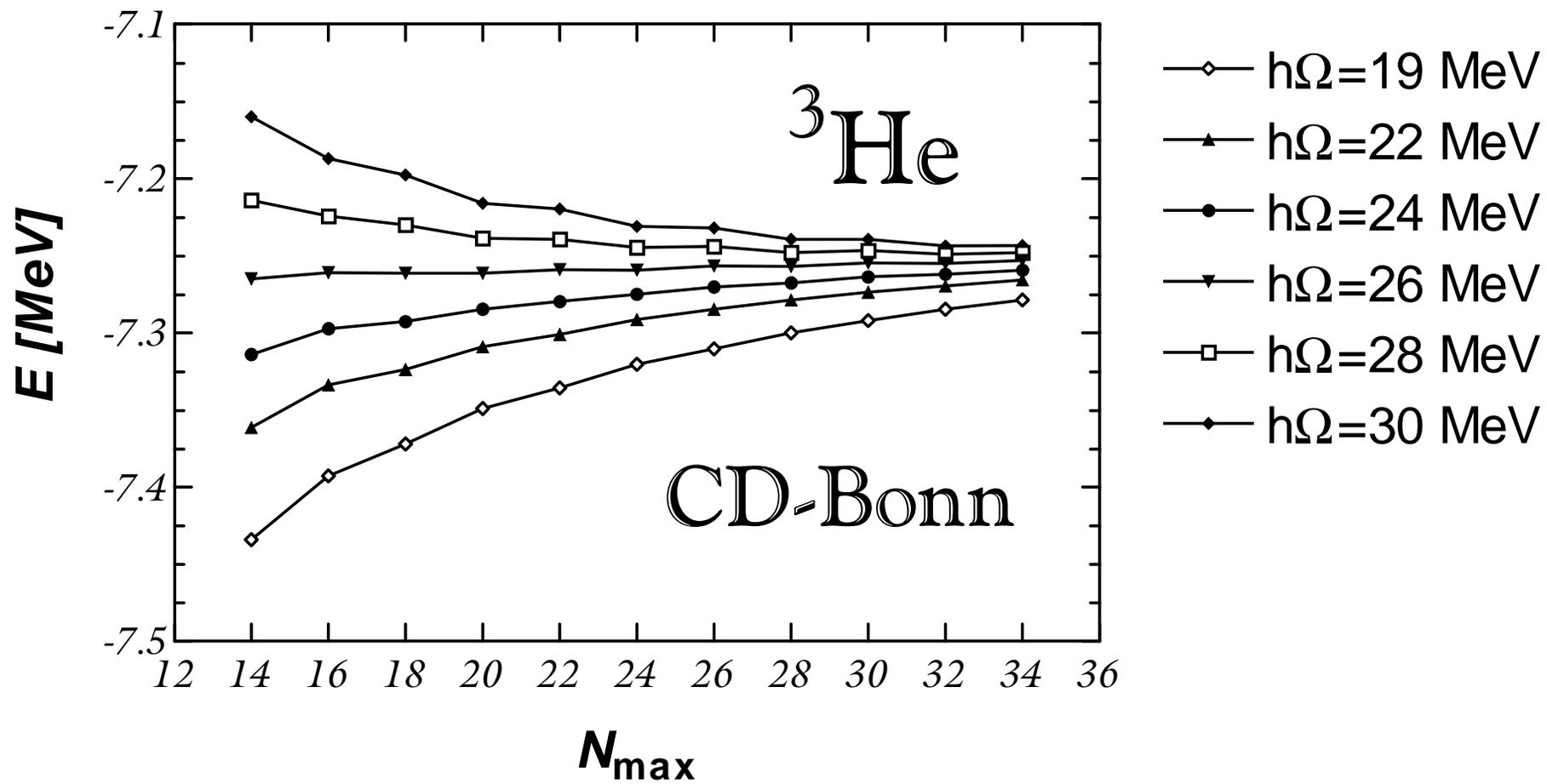
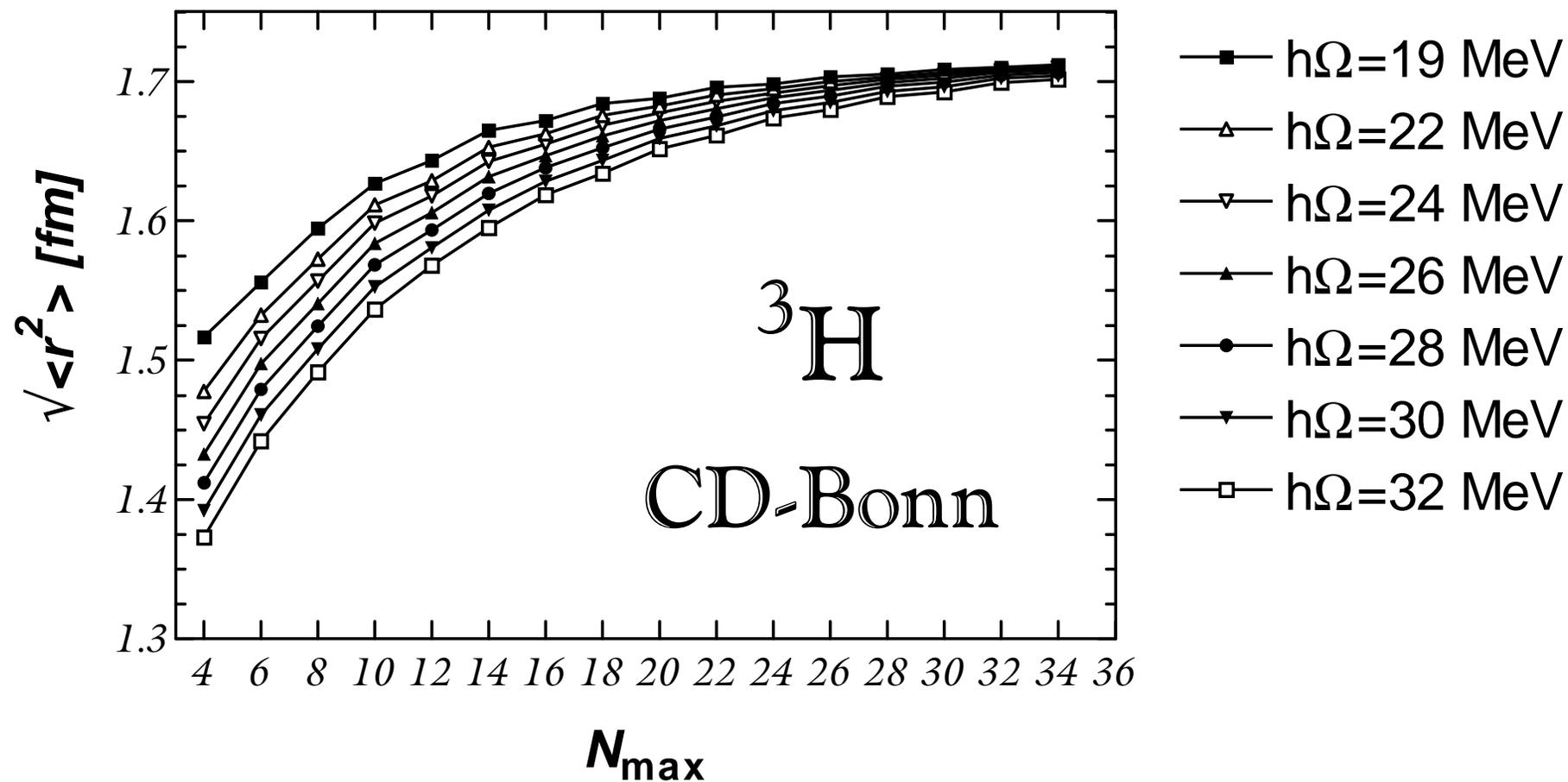
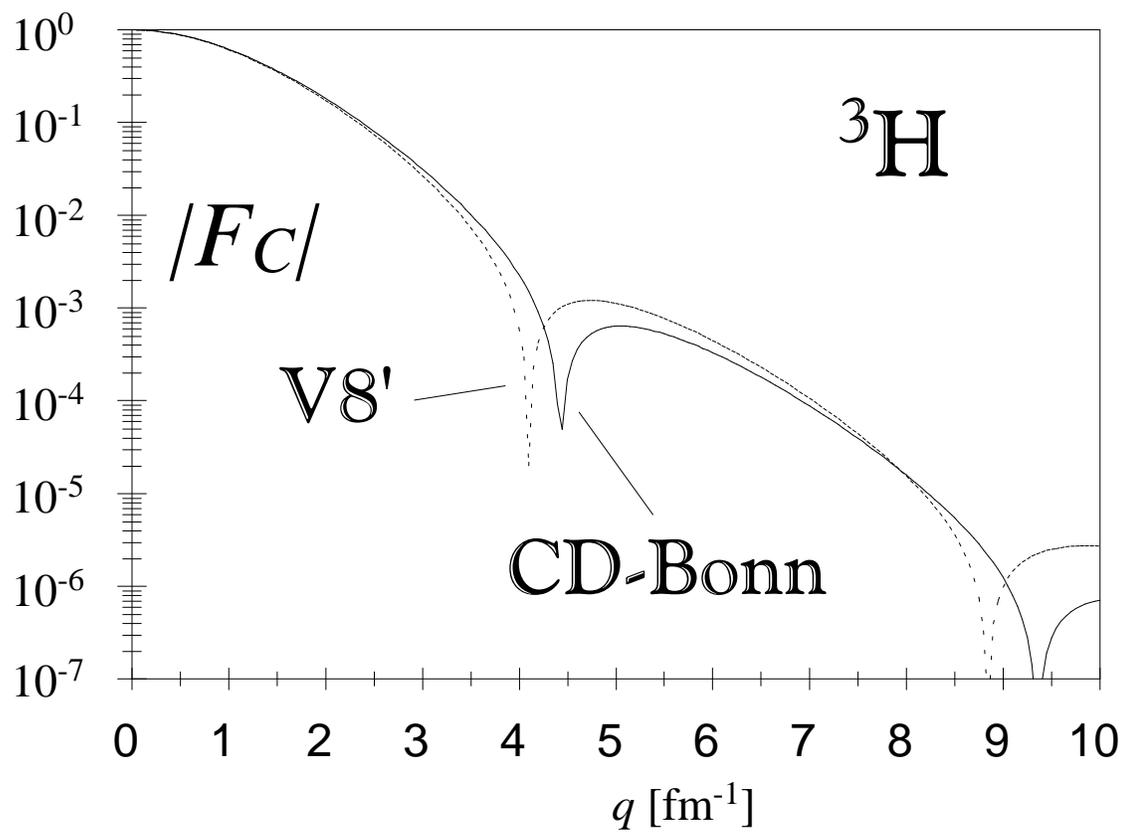
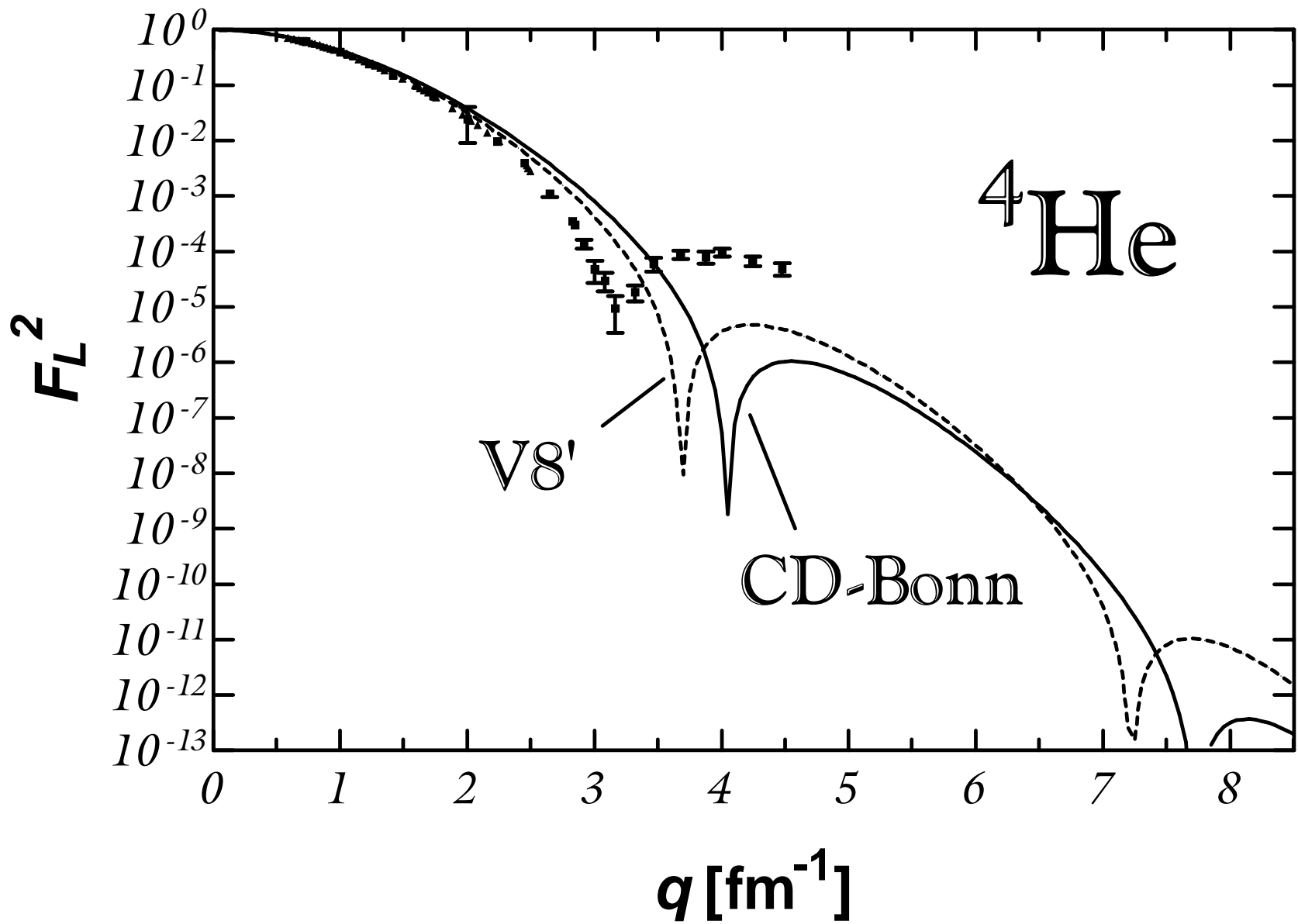
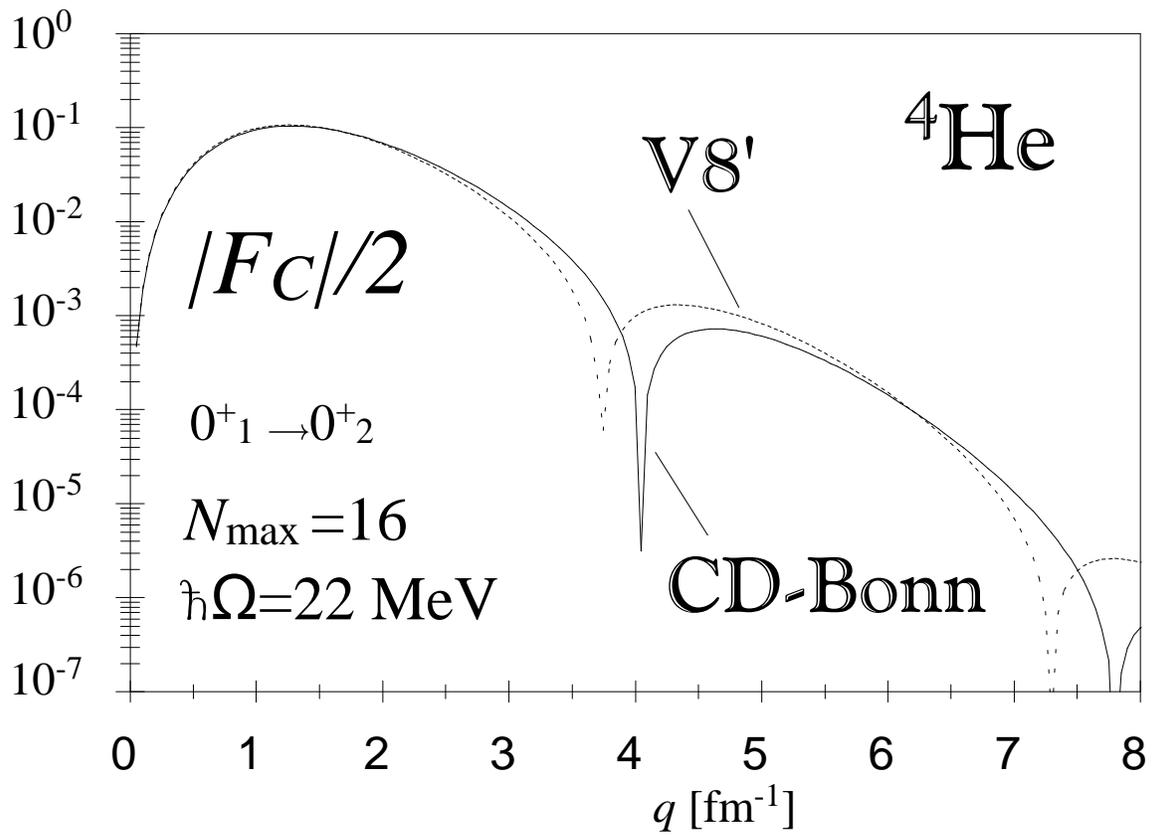
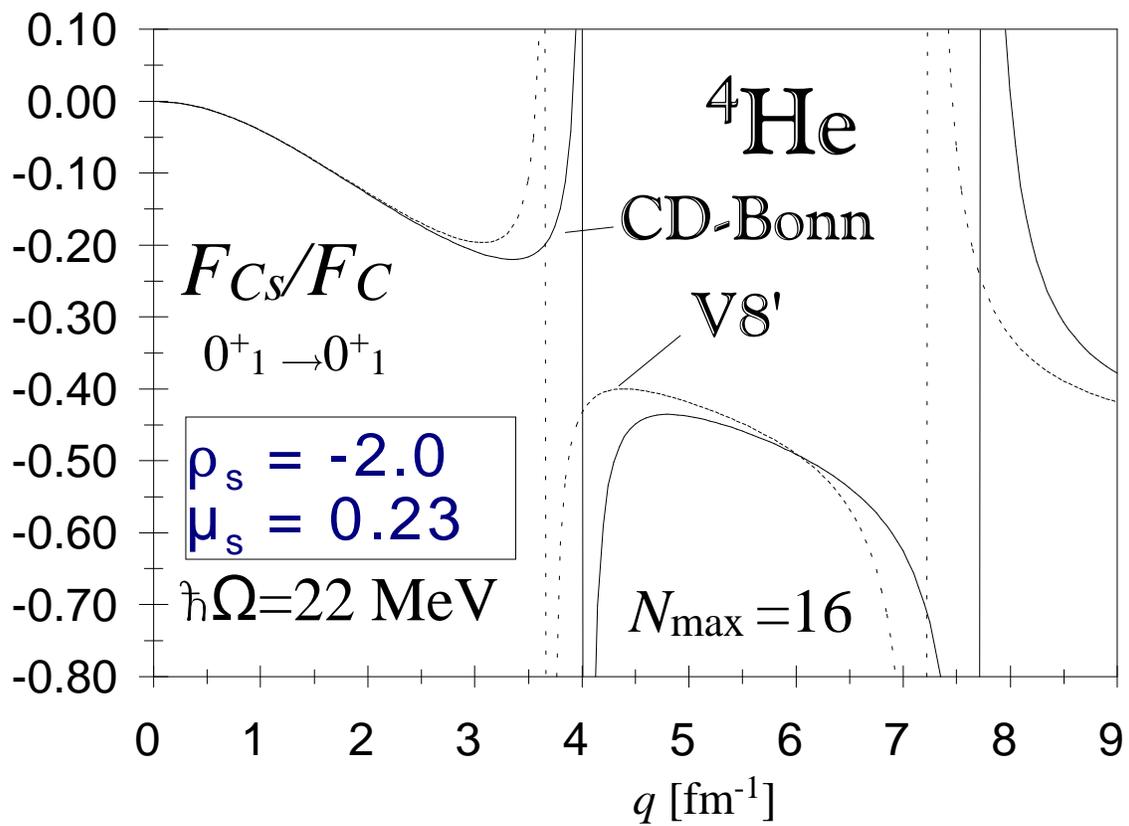
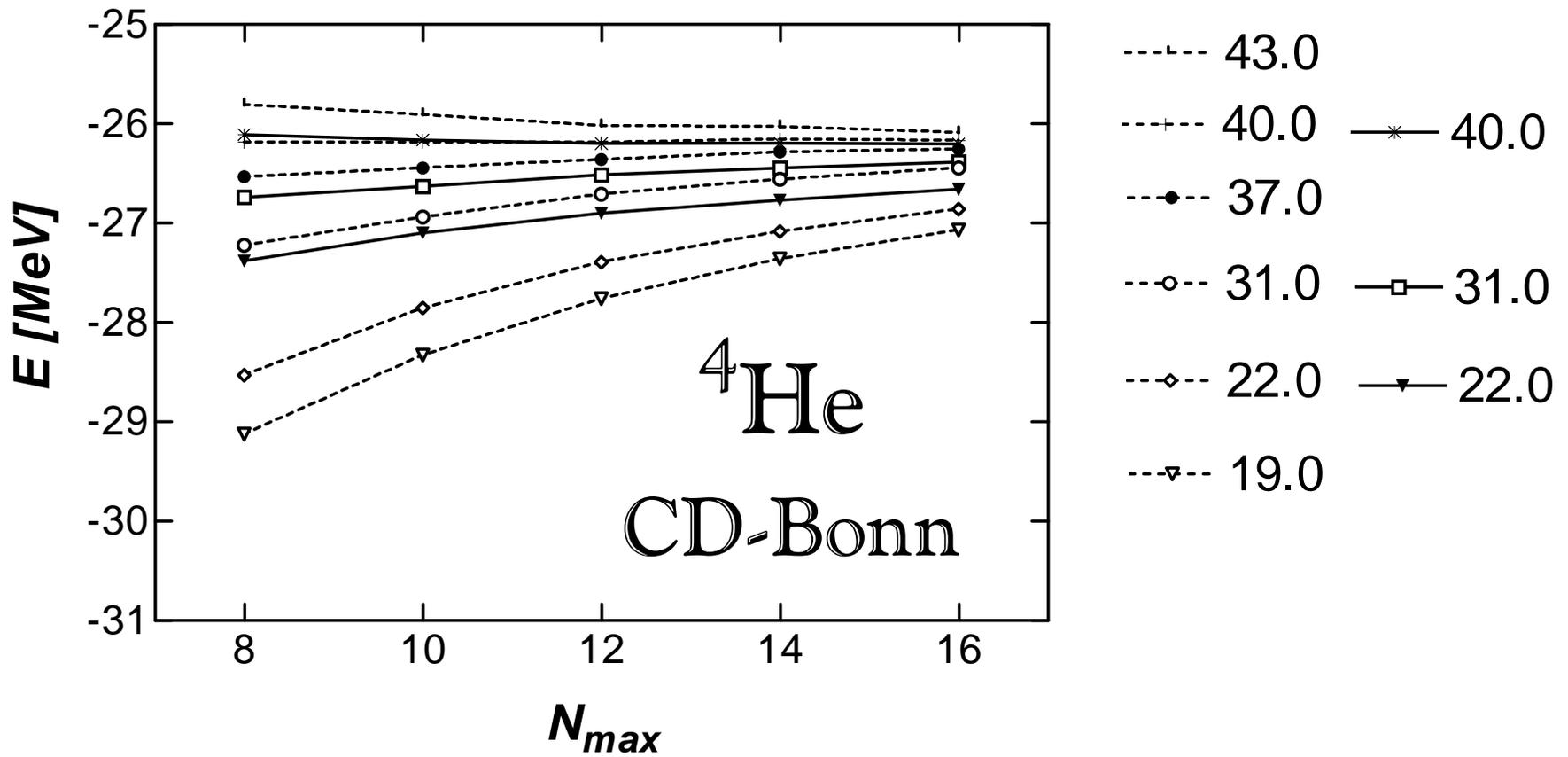
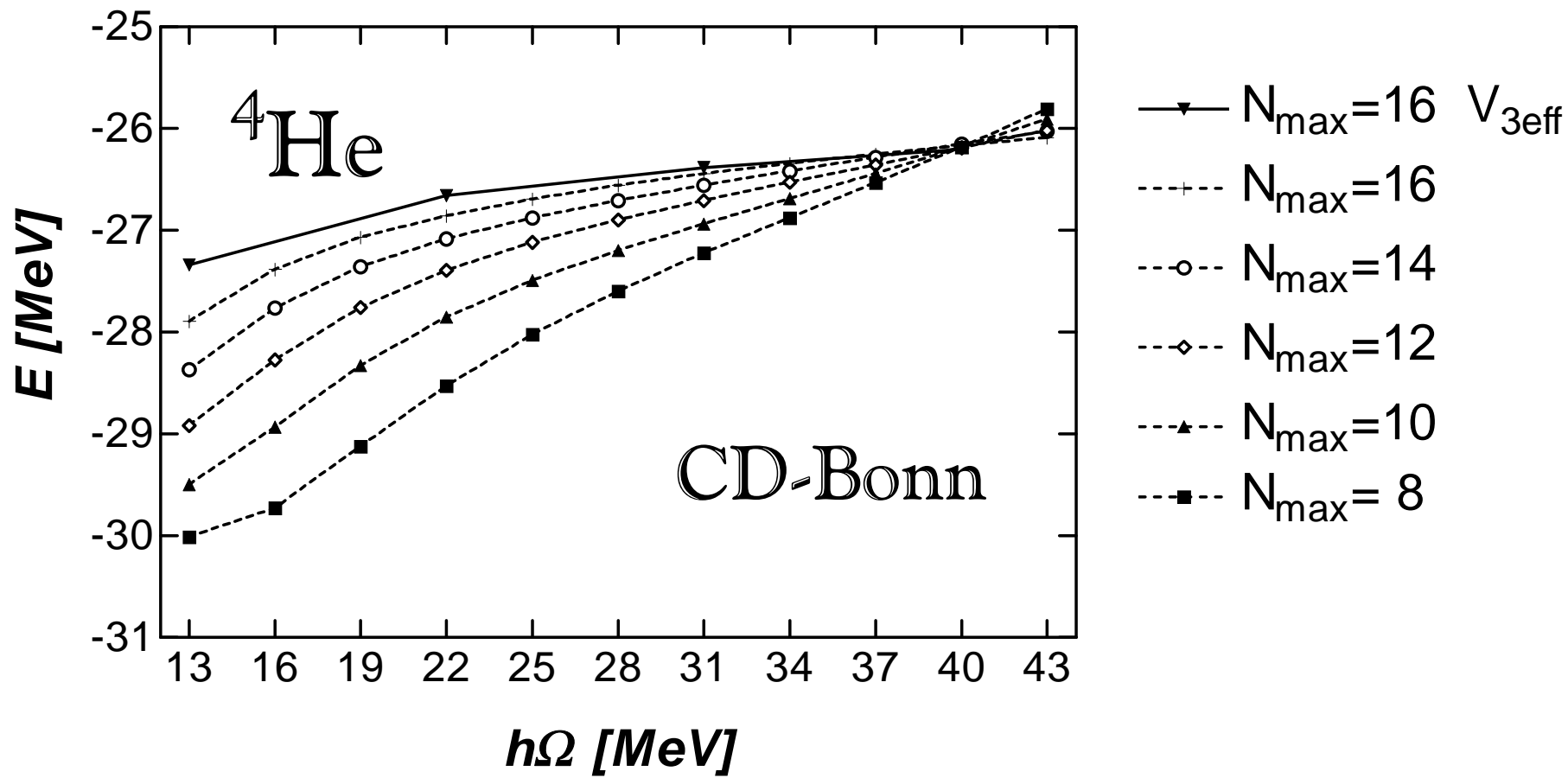
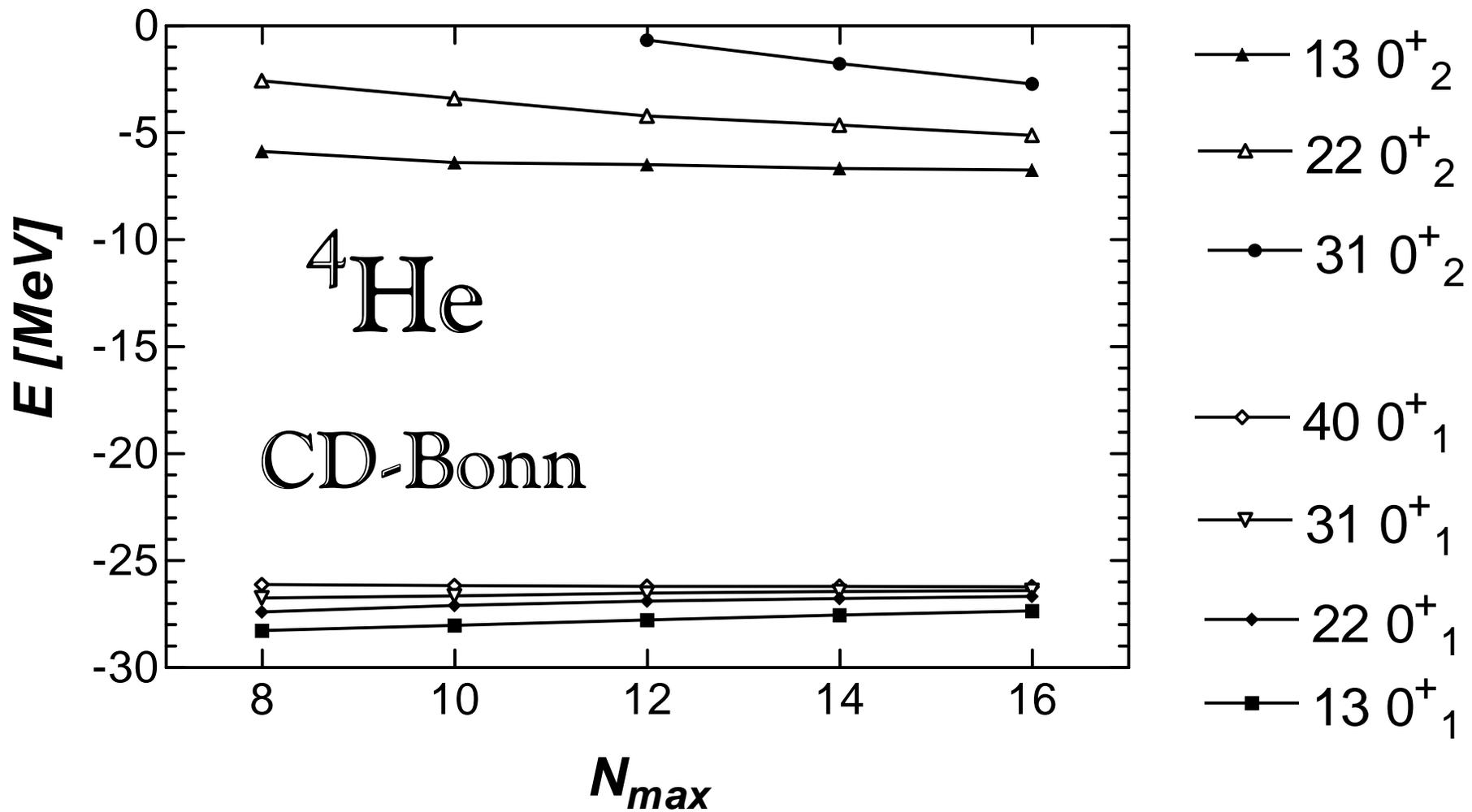
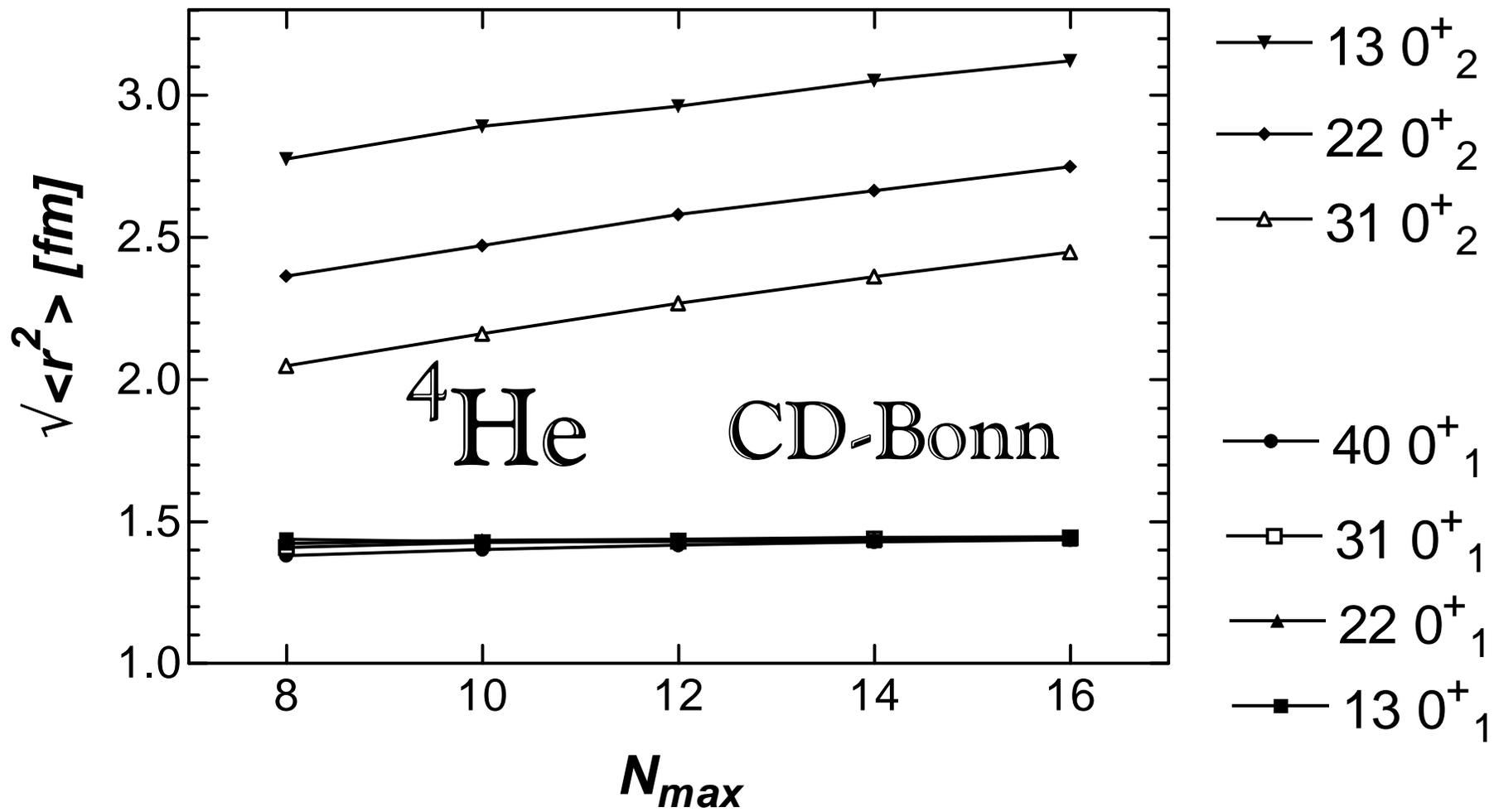
\end{document}